\RequirePackage[svgnames,dvipsnames]{xcolor}
\documentclass[openacc]{rsproca_new}

\usepackage{amsmath,amsthm}
\usepackage{amsfonts,amssymb,mathrsfs}
\usepackage{support-caption}
\usepackage[justification=raggedright, font={small,sf},labelfont={small,sf,bf}, singlelinecheck=off]{caption}
\usepackage{subcaption}
\usepackage{adjustbox}

\usepackage{color}
\usepackage[svgnames]{xcolor}
\definecolor{myurlcolor}{rgb}{0.6,0,0}
\definecolor{mycitecolor}{rgb}{0,0,0.9}
\definecolor{myrefcolor}{rgb}{0,0,0.9}
\definecolor{mygreen}{rgb}{0.2,1,0.6}

\definecolor{lblue}{rgb}{0,250,255}
\definecolor{llblue}{HTML}{a1ddde}
\definecolor{red}{rgb}{0.8,0,0}

\DeclareGraphicsExtensions{.pdf,.jpeg,.png}
\usepackage{csquotes}
\usepackage[all,2cell,cmtip]{xy}
\usepackage{tikz, tikz-cd}
\usetikzlibrary{backgrounds,circuits,circuits.ee.IEC,shapes,fit,matrix,shapes.geometric}

\tikzstyle{simple}=[-,line width=2.000]
\tikzstyle{arrow}=[-,postaction={decorate},decoration={markings,mark=at position .5 with {\arrow{>}}},line width=1.100]
\pgfdeclarelayer{edgelayer}
\pgfdeclarelayer{nodelayer}
\pgfsetlayers{edgelayer,nodelayer,main}

\tikzstyle{none}=[inner sep=-1pt]

\tikzstyle{species}=[circle, fill=yellow, draw=black, scale=1]
\tikzstyle{catalyst}=[circle, fill=yellow, draw=red, scale=1]
\tikzstyle{transition}=[rectangle, fill=lblue, draw=black, scale=1]
\tikzstyle{minitransition}=[rectangle, fill=lblue, draw=black, scale=0.8]
\tikzstyle{morphism}=[rectangle, fill=llblue, draw=black, scale=1]

\tikzstyle{empty}=[circle,fill=none, draw=none]
\tikzstyle{dot}=[circle,fill=black,draw=black, scale=.4]
\tikzstyle{bounding}=[circle,dashed, fill=none,draw=black, scale=9.00]
\tikzstyle{simple}=[-,draw=black,line width=1.000]
\tikzstyle{inarrow}=[->, >=stealth, shorten >=.03cm,line width=0.6]

\newcommand{\maps}{\colon}

\newcommand{\n}{\mathtt{n}}
\newcommand{\m}{\mathtt{m}}
\newcommand{\mplusn}{\mathtt{m+n}}

\newcommand{\N}{\mathbb{N}}

\newcommand{\define}[1]{{\bf \boldmath{#1}}}

\renewcommand{\S}{\mathcal{S}}

\newcommand{\Cat}{\mathbf{Cat}}

\newcommand{\Mon}{\mathbf{Mon}}

\newcommand{\Set}{\mathbf{Set}}
\newcommand{\Sem}{\mathbf{Sem}}

\newcommand{\X}{\mathtt{X}}
\newcommand{\Xone}{\mathtt{X_1}}
\newcommand{\Xn}{\mathtt{X_n}}
\newcommand{\Xitype}{\mathtt{X_i}}
\newcommand{\Xij}{\mathtt{X_{ij}}}
\newcommand{\Y}{\mathtt{Y}}
\newcommand{\Yi}{\mathtt{Y_i}}
\newcommand{\Ztype}{\mathtt{Z}}
\newcommand{\fop}{\mathtt{f}}
\newcommand{\gop}{\mathtt{g}}
\newcommand{\gopi}{\mathtt{g_i}}
\newcommand{\gopn}{\mathtt{g_n}}
\newcommand{\gopone}{\mathtt{g_1}}
\newcommand{\hoper}{\mathtt{h}}
\newcommand{\varphiOp}{\mathtt{f}}
\newcommand{\kappaOp}{\mathtt{g}}
\newcommand{\lambdaOp}{\mathtt{l}}
\newcommand{\tauOp}{\mathtt{t}}
\newcommand{\sigmaOp}{\mathtt{s}}
\newcommand{\alphaOp}{\mathtt{a}}
\newcommand{\opOp}{\mathtt{op}}
\newcommand{\opone}{\mathtt{op_1}}
\newcommand{\optwo}{\mathtt{op_2}}

\newcommand{\System}{\mathtt{System}}

\newcommand{\Subone}{\mathtt{Sub_1}}
\newcommand{\Subtwo}{\mathtt{Sub_2}}
\newcommand{\Suboneone}{\mathtt{Sub_{11}}}
\newcommand{\Subonetwo}{\mathtt{Sub_{12}}}

\newcommand{\Subtwoone}{\mathtt{Sub_{21}}}
\newcommand{\Subtwotwo}{\mathtt{Sub_{22}}}
\newcommand{\Subtwothree}{\mathtt{Sub_{23}}}

\newcommand{\Alg}{\mathsf{A}}

\newcommand\scalemath[2]{\scalebox{#1}{\mbox{\ensuremath{\displaystyle #2}}}}

\newcommand{\CUTTER}{\mathtt{Cut}}
\newcommand{\HELO}{\mathtt{Helo}}
\newcommand{\BOAT}{\mathtt{Boat}}
\newcommand{\QUAD}{\mathtt{QD}}
\newcommand{\FWS}{\mathtt{FW}}
\newcommand{\FWSAR}{\mathtt{FSAR}}
\newcommand{\FWUAV}{\mathtt{UAV}}
\newcommand{\PORT}{\mathtt{Port}}
\newcommand{\HH}{\mathtt{UH60}}

\newcommand{\HC}{\mathtt{HC130}}

\usepackage{mathtools}
\usepackage{multirow}

\newcommand{\<}{\langle}
\renewcommand{\>}{\rangle}

\newcommand{\id}{\mathsf{id}}

\newcommand{\SG}{\mathsf{SG}}
\newcommand{\MG}{\mathsf{MG}}

\newcommand{\PP}{\mathcal{P}}

\newcommand{\OO}{{\mathcal{O}}}
\newcommand{\NN}{\mathbb{N}}

\newcommand{\Bit}{\mathbf{Bit}}
\newcommand{\LG}{\mathsf{LG}}

\newcommand{\Interfer}{\mathtt{Intfr}}
\newcommand{\Chassis}{\mathtt{Chassis}}
\newcommand{\Optics}{\mathtt{Optics}}
\newcommand{\Bath}{\mathtt{Bath}}
\newcommand{\BoxTemp}{\mathtt{Box}}
\renewcommand{\Box}{\mathtt{Box}}

\newcommand{\Lab}{\mathtt{Lab}}

\newcommand{\TempSys}{\mathtt{TempSys}}
\newcommand{\LengthSys}{\mathtt{LengthSys}}
\newcommand{\Sensors}{\mathtt{Sensors}}
\newcommand{\Actuators}{\mathtt{Actuators}}
\newcommand{\LSIob}{\mathtt{LSI}}
\newcommand{\intens}{\mathtt{intensity}}
\newcommand{\drive}{\mathtt{drive}}

\newcommand{\heat}{\mathtt{heat}}
\newcommand{\laser}{\mathtt{laser}}
\newcommand{\water}{\mathtt{H_2O}}
\newcommand{\focus}{\mathtt{focus}}
\newcommand{\temp}{\mathtt{temp}}
\newcommand{\fringe}{\mathtt{fringe}}
\newcommand{\setPt}{\mathtt{setPt}}

\newcommand{\Req}{\mathsf{Req}}

\newcommand{\State}{\mathsf{State}}
\newcommand{\KPI}{\mathsf{KPI}}

\newcommand{\Val}{\mathsf{Val}}

\newcommand{\extr}{\mathsf{extr}}
\newcommand{\degC}{^\circ\textrm{C}}
\newcommand{\RR}{\mathbb{R}}

\newcommand{\Rel}{\mathbf{Rel}}

\newcommand{\cvect}[1]{
	\mbox{$\tiny\begin{pmatrix}#1\end{pmatrix}$}
}
\newcommand{\tdots}{{\scriptscriptstyle\ldots}}
\newcommand{\XVal}{\mathsf{XVal}}
\newcommand{\JVal}{\mathsf{JVal}}
\newcommand{\Context}{\mathsf{Context}}
\newcommand{\Ent}{\mathbf{Ent}}
\newcommand{\inst}{\mathsf{inst}}
\newcommand{\const}{\mathsf{const}}

\newcommand{\Traj}{\mathsf{Traj}}

\newcommand{\Rigid}{\mathsf{Rigid}}
\newcommand{\Lump}{\mathsf{Lump}}
\newcommand{\Dyn}{\mathsf{Dyn}}
\newcommand{\Impl}{\mathsf{Impl}}
\newcommand{\Net}{\mathcal{N}}
\newcommand{\Type}{\mathsf{Type}}
\newcommand{\Edlen}{\textsf{Edl\'{e}n}}

\newcommand{\WW}{\mathcal{W}}
\newcommand{\Prob}{\mathbf{Prob}}

\mathchardef\mhyphenhalf="2D
\newcommand{\mhyphen}{\mhyphenhalf\mhyphenhalf}

\newcommand{\beq}{\begin{equation}}
\newcommand{\eeq}{\end{equation}}
\newcommand{\beqa}{\begin{eqnarray}}
\newcommand{\eeqa}{\end{eqnarray}}

\jname{rspa}
\Journal{Proc R Soc A\ }

\begin{document}

\title[Operads for complex system design]{Operads for complex system design specification, analysis and synthesis}
\author{
John D. Foley$^{1}$, Spencer Breiner$^{2}$, Eswaran Subrahmanian$^{2,3}$  and John M. Dusel$^{1}$}

\address{$^{1}$Metron, Inc., 1818 Library St., Reston, VA, USA\\
$^{2}$US National Institute of Standards and Technology, Gaithersburg, MD, USA\\
$^{3}$Carnegie Mellon University, Pittsburgh, PA, USA}

\subject{mathematical modeling, artificial intelligence, category theory}

\keywords{complex systems, system design, automated reasoning, compositionality,  applied category theory, operads}

\corres{John D.\ Foley\\
\email{foley@metsci.com}}

\begin{abstract}
    As the complexity and heterogeneity of a system grows, the 
    challenge of specifying, documenting and synthesizing correct, machine-readable designs increases dramatically.
    Separation of the system into manageable parts is needed to support analysis 
    at various levels of granularity so that the system is maintainable and adaptable over its life cycle.
    In this paper, we argue that operads provide an effective knowledge representation to address these challenges.  Formal documentation of a syntactically correct design is built up during design synthesis, guided by semantic reasoning about design effectiveness.
    Throughout, the ability to decompose the system into parts and reconstitute the whole is maintained.
We describe recent progress in effective modeling under this paradigm 
and directions for future work to systematically address scalability challenges for complex system design.
\end{abstract}


\begin{fmtext}
\section{Introduction}
\label{sec:intro}

We solve complex problems by separating them into manageable parts \cite{Form, Simon1991}. Human designers do this intuitively, but details can quickly overwhelm intuition.  
Multiple aspects of a problem may lead to distinct decompositions and complementary models of a system--e.g.\ competing considerations for cyberphysical systems \cite{LeeReview, Cyberphysical}--or simulation of behavior at many levels of fidelity--e.g.\ in modeling and simulation \cite{MS}--leading to a spectrum of models which are challenging to align.
We argue that operads, formal tools developed to compose geometric and algebraic objects, are uniquely suited to separate complex systems into manageable parts 
and maintain alignment across complementary models.
\end{fmtext}

\maketitle
  \begin{figure}[!ht]
    \centering
\scalebox{0.85}{
\begin{tikzpicture}
	\node[draw,circle,inner sep=0pt,minimum size=5ex] (f) at (-1,0) {};
	\node (X1) at (-1,1) {};
	\node (X2) at (-2,-0.25) {};
	\draw (f) -- (X1);
	\draw (f) -- (X2);
	\node[draw,circle,inner sep=0pt,minimum size=3ex] (g1) at (-0.25,-0.9) {};
	\node (Z1) at (-1.5,-1.5) {};
	\node (Z2) at (0,-1.5) {};
	\draw (g1) -- (Z1);
	\draw (g1) -- (Z2);
	\node[draw,circle,inner sep=0pt,minimum size=2ex] (g2) at (0.6,-0.3) {};
	\node (Y) at (1, -0.8) {};
	\draw (g2) -- (Y);
	\draw (f) -- (g1);
	\draw (g1) -- (g2);
	\draw (f) -- (g2);
	%
	%
	\node[style=none] (G) at (-2.2,-1.05) {(1)};
	\node[style=none] (Cen) at (-0.3, -0.4) {};
	\node (Top) at (-0.3, 1) {};
	\node (Lef) at (-2,-1) {};
	\node (Rig) at (1,-1.2) {};
	\draw[thick, ForestGreen] (Cen) -- (Top);
		\draw[thick, ForestGreen] (Cen) -- (Lef);
			\draw[thick, ForestGreen] (Cen) -- (Rig);
	\node[draw,inner sep=0pt,minimum size=5ex] (Af) at (3,1) {};
	\node (AX1) at (3,2) {};
	\node (AX2) at (2,0.75) {};
79

	\draw (Af) -- (AX1);
	\draw (Af) -- (AX2);
	\node[draw,inner sep=0pt,minimum size=3ex] (Ag1) at (3.75,0.1) {};
	\node (AZ1) at (2.5,-0.5) {};
	\node (AZ2) at (4,-0.5) {};
	\draw (Ag1) -- (AZ1);
	\draw (Ag1) -- (AZ2);
	\node[draw,inner sep=0pt,minimum size=2ex] (Ag2) at (4.6,0.7) {};
	\node (AY) at (5, 0.2) {};
	\draw (Ag2) -- (AY);
	\draw (Af) -- (Ag1);
	\draw (Ag1) -- (Ag2);
	\draw (Af) -- (Ag2);
	\node[draw, diamond, inner sep=0pt,minimum size=5ex] (Bf) at (2.5,-2) {};
	\node (BX1) at (2.5,-1) {};
	\node (BX2) at (1.5,-2.25) {};
	\draw (Bf) -- (BX1);
	\draw (Bf) -- (BX2);
	\node[draw, diamond, inner sep=0pt,minimum size=3ex] (Bg1) at (3.25,-2.9) {};
	\node (BZ1) at (2,-3.5) {};
	\node (BZ2) at (3.5,-3.5) {};
	\draw (Bg1) -- (BZ1);
	\draw (Bg1) -- (BZ2);
	\node[draw,diamond,inner sep=0pt,minimum size=2ex] (Bg2) at (4.1,-2.3) {};
	\node (BY) at (4.5, -2.8) {};
	\draw (Bg2) -- (BY);
	\draw (Bf) -- (Bg1);
	\draw (Bg1) -- (Bg2);
	\draw (Bf) -- (Bg2);
	%
	\node[style=none] (R) at (0.8,1.6) {(2)};
	\node[style=none] (CenMain) at (1.5, -0.75) {};
 	\node (TopMain) at (1.1,2) {};
 	\node (BotMain) at (0.3,-3.4) {};
 	\draw[very thick, red] (CenMain) -- (TopMain);
 		\draw[very thick, red] (CenMain) -- (BotMain);
 	\node (O1) at (0.6, 0.5) {};
 	\node (A1) at (2, 0.8) {};
 	\draw[->, thick] (O1) to (A1);
 	\node (O2) at (0.4, -1.5) {};
 	\node (B1) at (1.6, -2) {};
 	\draw[->, thick] (O2) to (B1);
	%
	\node[style=none] (R) at (4.65,-0.85) {(3)};
 	\node (RigMain) at (5,-1.2) {};
 		\draw[very thick, blue] (CenMain) -- (RigMain);
 	\node (A2) at (3.4, -0.5) {};
 	\node (B2) at (3.2, -1.6) {};
 	\draw[-implies, thick,
		double equal sign distance] (A2) to (B2);
	\node[align=center] (SynLab) at (-4.5,1.5) {\large Syntax};
	\node[draw,circle,inner sep=0pt,minimum size=1ex] (Type) at (-5.5,0) {};
	\node (TypeUp) at (-5.5, 0.25) {};
	\node (TypeRight) at (-5.25, 0) {};
	\node (TypeDL) at (-5.72, -0.2) {};
	\draw (Type) -- (TypeUp);
	\draw (Type) -- (TypeRight);
	\draw (Type) -- (TypeDL);
	\node[align=center] (TypeLab) at (-4,0) {Abstract\\system designs};
	\node[draw,circle,inner sep=0pt,minimum size=1ex] (T1) at (-5.7,-0.7) {};
	\node (T1Up) at (-5.7, -0.45) {};
	\node (T1Right) at (-5.45, -0.7) {};
	\node (T1DL) at (-5.92, -0.9) {};
	\draw (T1) -- (T1Up);
	\draw (T1) -- (T1Right);
	\draw (T1) -- (T1DL);
    \node[draw,circle,inner sep=0pt,minimum size=1ex] (T2) at (-5.3,-0.7) {};
	\node (T2Down) at (-5.3, -0.95) {};
	\node (T2Left) at (-5.55, -0.7) {};
	\draw (T2) -- (T2Down);
	\draw (T2) -- (T2Left);
   \node [rotate=-90] at (-5.5, -1) {$\mapsto$};
	\node[align=center] (CompLab) at (-4,-1) {Composing\\designs};
	\node[draw,circle,inner sep=0pt,minimum size=1ex] (Tout1) at (-5.7,-1.3) {};
	\node (Tout1Up) at (-5.7, -1.05) {};
	\node (Tout1DL) at (-5.92, -1.5) {};
	\draw (Tout1) -- (Tout1Up);
	\draw (Tout1) -- (Tout1DL);
    \node[draw,circle,inner sep=0pt,minimum size=1ex] (Tout2) at (-5.3,-1.3) {};
    \draw (Tout1) -- (Tout2);
	\node (Tout2Down) at (-5.3, -1.55) {};
	\draw (Tout2) -- (Tout2Down);
	\node[align=center] (SynLab) at (7,1.5) {\large Semantics};
	\node[draw,inner sep=0pt,minimum size=1ex] (AA) at (5.8,0) {};
	\node (AAUp) at (5.8, 0.25) {};
	\node (AARight) at (6.05, 0) {};
	\node (AADL) at (5.58, -0.2) {};
	\draw (AA) -- (AAUp);
	\draw (AA) -- (AARight);
	\draw (AA) -- (AADL);
	\node[draw,diamond,inner sep=0pt,minimum size=1ex] (BB) at (6.2,0) {};
	\node (BBUp) at (6.2, 0.25) {};
	\node (BBRight) at (6.45, 0) {};
	\node (BBDL) at (5.98, -0.2) {};
	\draw (BB) -- (BBUp);
	\draw (BB) -- (BBRight);
	\draw (BB) -- (BBDL);
	\node[align=center] (AlgLab) at (7.5,0) {Computational\\models};
	\node[draw,inner sep=0pt,minimum size=1ex] (AA1) at (5.8,-0.7) {};
	\node (AA1Up) at (5.8, -0.45) {};
	\node (AA1Right) at (6.05, -0.7) {};
	\node (AA1DL) at (5.58, -0.9) {};
	\draw (AA1) -- (AA1Up);
	\draw (AA1) -- (AA1Right);
	\draw (AA1) -- (AA1DL);
    \node[draw,inner sep=0pt,minimum size=1ex] (AA2) at (6.2,-0.7) {};
	\node (AA2Down) at (6.2, -0.95) {};
	\node (AA2Left) at (5.95, -0.7) {};
	\draw (AA2) -- (AA2Down);
	\draw (AA2) -- (AA2Left);
	\node [rotate=-90] at (6, -1) {$\mapsto$};
	\node[align=center] (AlgComp) at (7.5,-1) {Composing\\models};
	\node[draw,inner sep=0pt,minimum size=1ex] (AAout1) at (5.8,-1.3) {};
	\node (AAout1Up) at (5.8, -1.05) {};
	\node (AAout1DL) at (5.58, -1.5) {};
	\draw (AAout1) -- (AAout1Up);
	\draw (AAout1) -- (AAout1DL);
    \node[draw,inner sep=0pt,minimum size=1ex] (AAout2) at (6.2,-1.3) {};
    \draw (AAout1) -- (AAout2);
	\node (AAout2Down) at (6.2, -1.55) {};
	\draw (AAout2) -- (AAout2Down);
\end{tikzpicture}
}
\vspace*{-10pt}
    \caption{Separating concerns with operads: (1) Composition separates subsystem designs (green boundaries {\bf \textcolor{ForestGreen}{--}}); (2) Functorial semantics separate abstract systems from computational model instances (red boundaries {\bf \textcolor{red}{--}}); (3) Natural transformations separate and align (blue boundary {\bf \textcolor{blue}{--}}) complementary  models ($\square$, \protect\scalebox{1.5}{$\diamond$}).}
    \label{fig:my_label}
    \vspace*{-8pt}
\end{figure}
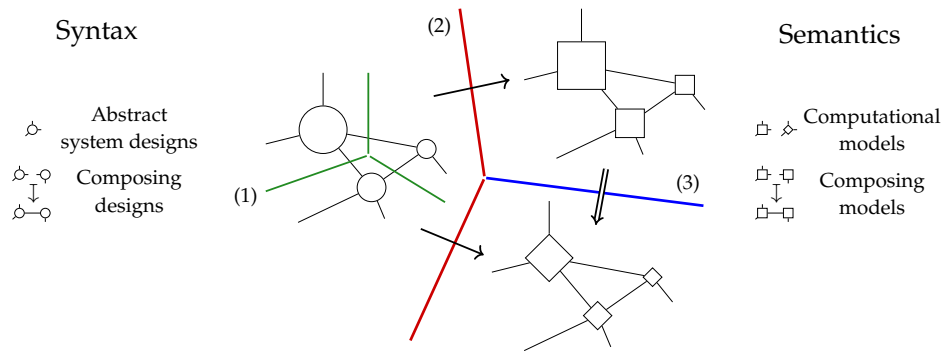

Operads provide three ways to separate concerns for complex systems: (1) designs for subsystems are separated into composable modules; (2) syntactic designs to compose systems are separated from the semantic data that model them; and (3) separate semantic models can be aligned to evaluate systems in different ways.
The three relationships are illustrated in Figure~\ref{fig:my_label}.  

Hierarchical decomposition (Fig.~\ref{fig:my_label}, {\bf \textcolor{ForestGreen}{--}}) is nothing new.  Both products and processes are broken down to solve problems from design to daily maintenance.  Operads provide a precise language  
to manage complex modeling details that the intuitive--and highly beneficial--practice of decomposition uncovers, e.g., managing multiple, complementary decompositions and models.    

Operads separate the syntax to compose subsystems from the semantic data modeling them (Fig.~\ref{fig:my_label}, {\bf \textcolor{red}{--}}).  Syntax consists of abstract ``operations'' to design the parts and architecture of a system.  Semantics define how to interpret and evaluate these abstract blueprints.  
Operad syntax is typically lightweight and declarative. Operations can often be represented both graphically and algebraically (Fig. \ref{fig:decomps}), formalizing intuitive design diagrams. Operad semantics model specific aspects of a system and can range from fast to computationally expensive.

The most powerful way operads separate is by aligning complementary models  while maintaining  compatibility with system decompositions (Fig.~\ref{fig:my_label}, {\bf \textcolor{blue}{--}}).
Reconciling complementary models is a persistent and pervasive issue across domains \cite{ModelInt, RelateInt, SimBot, MultiOne, LeeReview, SemReconcil, MathMod, MultiBio}.
Historically, Eilenberg \& Mac Lane \cite{EM} invented \emph{natural transformations} to align computational models of topological spaces.
Operads use natural transformations to align hierarchical decompositions, which is particularly well-suited to system design. 

This paper articulates a uniform and systematic foundation for system design and analysis. In essence, the syntax of an operad defines \emph{what can be} put together, which is a prerequisite to decide \emph{what should be} put together.  Interfaces define which designs are \emph{syntactically} feasible, but key \emph{semantic} information must be expressed to evaluate candidate designs. Formulating system models within operad theory enforces the intellectual hygiene required to make sure that different concerns stay separated while working together to solve complex design problems.
 
We note five strengths of this foundation that result from the three ways operads separate a complex problem and key sections of the paper that provide illustrations.

{\bf Expressive, unifying meta-language.}
A meta- or multi-modeling \cite{MultiMod} language is needed to express and relate multiple representations.
The key feature of operad-based meta-modeling is its focus on coherent mappings between models (Fig.~\ref{fig:my_label}, {\bf \textcolor{red}{--}}, {\bf \textcolor{blue}{--}}), as opposed to a universal modeling context, like UML, OWL, etc., which is inevitably under or over expressive for particular use cases.  
Unification allows complementary models to work in concert, as we see in Sec.\ \ref{sec:anal} for function and control. 
Network operads---originally developed to design systems---were applied to task behavior.  
This power to unify and express becomes especially important when reasoning across domains with largely independent modeling histories; compare, e.g., \cite{Cyberphysical}. 
(\ref{sec:operads}\ref{subsec:API_view}, \ref{sec:operads}\ref{subsec:mapmaker}, \ref{sec:examples}, \ref{sec:cookbook}\ref{subsec:cooknetmod}, \ref{sec:anal})

{\bf Minimal data needed for specification.}
 Data needed to set up each representation of a problem is minimal in two ways: (1) any framework must provide similar, generative data; and (2) each level only needs to specify data relevant to that level.
Each representation is self-sufficient and can be crafted to efficiently address a limited portion of the full problem.  
The modeler  can pick and choose relevant representations and extend the meta-model as needed.  (\ref{sec:cookbook}, \ref{sec:auto}\ref{subsec:autoTask})

{\bf Efficient exploration of formal correct designs.} An operad precisely defines how to iteratively construct designs or adapt designs by substituting different subsystems.
Constructing syntactically 
invalid designs is not possible, restricting the relevant design space, and correctness is preserved when moving across models.
 Semantic reasoning guides synthesis--potentially at several levels of detail. 
This facilitates lazy evaluation: first syntactic correctness is guaranteed, then multitudes of coarse models are explored before committing to later, more expensive evaluations. 
The basic moves of iteration, substitution, and moving across levels constitute a rich framework for exploration. 
We obtain not only an effective design but also formal documentation of the models which justify this choice.
(\ref{sec:operads}\ref{subsec:API_view}--\ref{subsec:equate_view}, \ref{sec:auto}, \ref{sec:sepcon}\ref{subsec:opProg})

{\bf Separates representation from exploitation.}
Operads and algebras provide structure and representation for a problem.  
Exploitation of this structure and representation is a separate concern.
As Herbert Simon noted during his Nobel Prize speech \cite{Simon}:
``\ldots decision makers can satisfice either by finding optimum solutions for a simplified world, or by finding satisfactory solutions for a more realistic world.''
This is an either-or proposition for a simple representation. By laying the problem across multiple semantic models,
useful data structures for each model--e.g.\ logical, evolutionary or planning frameworks--can be exploited by algorithms that draw on operad-based iteration and substitution.
 (\ref{sec:auto}, \ref{sec:sepcon}\ref{subsec:opProg})

{\bf Hierarchical analysis and synthesis.} Operads naturally capture options for the hierarchical decomposition of a system, either within a semantic model to break up large scale problems or across
models to gradually increase modeling fidelity.
(\ref{sec:operads}\ref{subsec:tree_view},  \ref{sec:anal},\ref{sec:auto}\ref{subsec:autoOther}, \ref{sec:sepcon}\ref{subsec:autolessons})

\subsection{Contribution to design literature}
\label{subsec:accomplishments}

There are well-known examples of the successful exploitation of separation. 
For instance, electronic design automation (EDA) has had decades of success leveraging hierarchical separation of systems and components to achieve  very large scale integration (VLSI) of complex electronic systems \cite{VLSI, NewParadigmVLSI, EDAVLSI}. 
 We do not argue that operads are needed for extremely modular domains. 
 Instead, operads may help broaden the base of domains that benefit from separation and provide a means to integrate and unify treatment across domains. 
On the other hand, for highly integral domains the ability to separate in practically useful ways may be limited \cite{SosaEtAl, Whitney}.
The recent applications we present help illustrate where operads may prove useful in the near and longer term; see \ref{sec:sepcon}\ref{subsec:advancementsAndLimits} for further discussion.

Compared to related literature, this article is application driven and outward focused.
Interest in applying operads and category theory to systems engineering has surged \cite{CompAI, FongThesis,  CTcouple, MR, AssemPlan, SpivakTan} 
as part of a broader wave applying category theory to design databases, software, proteins, etc.\ \cite{BSW,  DM, Seven, Matriarch,  CatSci, SpiProtein, SpivakKent}.
While much of loc.\ cit.\ matches applications to existing theoretical tools, the present article describes recent \emph{application driven} advancements and overviews \emph{specific methods} developed to address challenges presented by domain problems.  We introduce operads for design to a general scientific audience by explaining what the operads do relative to broadly applied techniques and how specific domain problems are modeled. Research directions are presented with an eye towards opening up interdisciplinary partnerships and continuing application driven investigations to build on recent insights.    

\subsection{Organization of the paper}
The present article captures an intermediate stage of technical maturity: operad-based design has shown its practicality by lowering barriers of entry for applied practitioners and demonstrating applied examples across many domains. However, it has not realized its full potential as an applied meta-language.  
Much of this recent progress is not focused solely on the analytic power of operads to separate concerns.  
Significant progress on explicit specification of domain models and techniques to automatically synthesize designs from basic building blocks has been made.
 Illustrative use cases and successful applications for
 design specification, analysis and synthesis organize the exposition.
		
Section \ref{sec:operads} introduces operads for design by analogy to other modeling approaches.  
Our main examples are introduced in Section \ref{sec:examples}.
Section \ref{sec:cookbook} describes how concrete domains can be specified with minimal combinatorial data,  lowering barriers to apply operads. 
 Section \ref{sec:anal} concerns analysis of a system with operads.
Automated synthesis is discussed in Section \ref{sec:auto}.
 Future research directions are outlined in  Section \ref{sec:sepcon}, which includes a list of open problems.
 
\begin{figure}
\centering
\begin{tikzpicture}
	\node[draw,ellipse,align=center] (Spec) at (0,-0.6) {Ex.\ 
	\ref{sec:examples}\ref{subsec:exSailboat}\\
	Specification\\Section \ref{sec:cookbook}};
	\node[draw,ellipse,align=center] (Ana) at (-5,-2.5) 
	{Ex.\ \ref{sec:examples}\ref{subsec:wireOPIntro}\\
	$\;$$\;$$\;$Analysis$\;$$\;$$\;$\\ Section \ref{sec:anal}};
	\node[draw,ellipse,align=center] (Syn) at (5,-2.5) {Ex.\ \ref{sec:examples}\ref{subsec:netOPIntro}\\
	$\;$$\;$Synthesis$\;$$\;$\\Section \ref{sec:auto}};
	\draw (Syn) to node[midway, above,  sloped] {Generators} (Spec);
	\draw (Spec) to node[midway, above,  sloped] {Compositionality} (Ana);
	\draw (Syn) to node[midway, above,  sloped] {Scalability} (Ana);
\end{tikzpicture}
		\vspace*{-4pt}
		\caption{Organization of the paper around applied examples introduced in Sec.\ \ref{sec:examples}.}
    \label{fig:org}
    \vspace*{-15pt}
\end{figure}
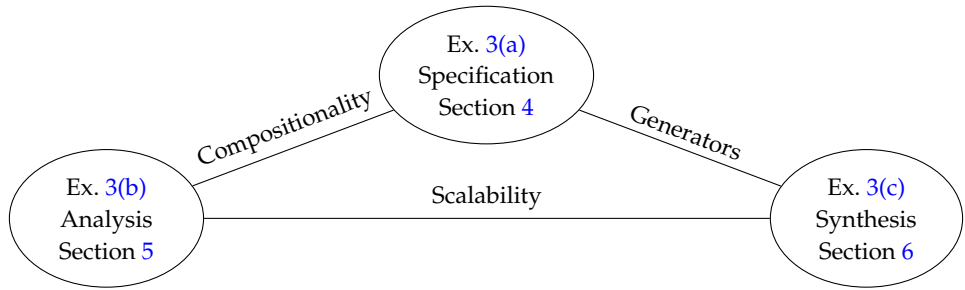

{\bf Notations.}  Throughout, we maintain the following notional conventions for: \vspace*{-7pt}
\begin{itemize}
    \item syntax operads (Fig.\ \ref{fig:my_label}, left), capitalized calligraphy: $\mathcal{O}$
    \item types (Fig.\ \ref{fig:my_label}, edges on left), capitalized teletype: $\X, \Y, \Ztype, \ldots$
    \item operations (Fig.\ \ref{fig:my_label}, nodes on left), uncapitalized teletype: $\fop, \gop, \hoper, \ldots$
    \item semantic contexts (Fig.\ \ref{fig:my_label}, right), capitalized bold: $\Sem, \Set, \Rel, \dots$
    \item functors from syntax to semantics  (Fig.\ \ref{fig:my_label}, arrows across red), capitalized sans serif: $\mathsf{Model} \maps \OO \to \Sem$;
    \item alignment of semantic models via natural transformations (Fig.\ \ref{fig:my_label}, double arrow across blue), uncapitalized sans serif: $\mathsf{align} \maps \mathsf{Model_1} \Rightarrow \mathsf{Model_2}$;
\end{itemize} \vspace*{-5pt}

\section{Applying operads to design}
\label{sec:operads}

We introduce operads by an analogy, explaining what an operad is and motivating its usefulness for systems modeling and analysis. The theory \cite{Leinster, MSS, Yau} pulls together many different intuitions. Here we highlight four analogies or `views' of an operad: hierarchical representations (tree view), strongly-typed programming  languages (API \footnote{Application Programming Interface}
 view), algebraic equations (equational view) and system cartography (map-maker's view).  Each view motivates operad concepts; see Table \ref{table:views}.

The paradigm of this paper is based on a typed operad, also known as a `colored operad' \cite{Yau} or `symmetric multicategory' \cite[2.2.21]{Leinster}.  
A typed \define{operad} $\OO$ has:\vspace*{-7pt}
\begin{itemize}
    \item A set $T$ of \define{types}.
    \item Sets of \define{operations} $\OO(\Xone,\ldots,\Xn ; \Y)$ where $\Xitype, \Y \in T$ and we write $\fop \maps \< \Xitype\> \to \Y$ to indicate that  $\fop\in\OO(\Xone,\ldots,\Xn ; \Y)$.
    \item A specific way to \define{compose} any operation $ \fop \maps \< \Yi\> \to \Ztype$
    with $\gopi \maps \<\Xij\> \to \Yi$ whose output types match the inputs of $f$ 
    to obtain a composite $\fop \circ (\gopone,\dots,\gopn) = \hoper \maps \<\Xij\> \to  \Ztype$.
    \end{itemize}
    These data are subject to rules \cite[11.2]{Yau} governing permutation of arguments and assuring that iterative composition is coherent, analogous to associativity for ordinary categories \cite[I]{MacLane}.  
	\vspace*{-7pt}
\begin{table}[!ht]
\centering	
\caption{The theory of operads draws on many familiar ideas, establishing a dictionary between contexts.}
\begin{tabular}{c|cccc}
	       \textbf{Operads} & \textbf{Tree} & \textbf{API} & \textbf{Equational} & \textbf{Systems} \\
	    \hline
		Types & Edges & Data types & Variables &  Boundaries \\
		Operations & Nodes & Methods & Operators & Architectures \\
		Composites & Trees & Scripts & Evaluation & Nesting \\
		Algebras & Labels & Implementations & Values & Models  \\
	\end{tabular}
	\label{table:views}
	\vspace*{-5pt}
\end{table}
	
\subsection{The tree view}
\label{subsec:tree_view}

Hierarchies are everywhere, from scientific and engineered systems to government, business and everyday life; they help to decompose complex problems into more manageable pieces.  The fundamental constituent  of an operad, called an \emph{operation}, represents a single step in a hierarchical decomposition. We can think of this as a single branching in a labeled tree, e.g.:  
\[
\scalebox{0.85}{\begin{tikzpicture}
	\node[draw,circle] (f) at (0,0) {$\opOp$};
	\node (X) at (0,1) {$\System$};
	\node (Y1) at (-.5,-1) {$\Subone$};
	\node (Y2) at (.5,-1) {$\Subtwo$};
	\draw (f) -- (X);
	\draw (f) -- (Y1);
	\draw (f) -- (Y2);
\end{tikzpicture}}\]
 Formally, this represents an element $\opOp  \in \OO(\Subone, \Subtwo ; \System)$. More generally, we can form new operations---trees---by \emph{composition}. 
Given further refinements for the two subsystems $\Subone$ and $\Subtwo$, by $\opone$ and $\optwo$, respectively, we have three composites:
\begin{equation}
\begin{array}{ccc}
\scalebox{0.85}{
\begin{tikzpicture}
	\node[draw,circle] (f) at (0,0) {$\opOp$};
	\node[draw,circle,inner sep=0pt,minimum size=5ex] (g1) at (-.75,-1) {$\opone$};
	\node (X) at (0,1) {$\System$};
	\node (Z1) at (-1.5,-2) {$\Suboneone$};
	\node (Z2) at (-.25,-2) {$\Subonetwo$};
	\node (Y2) at (1,-2) {$\Subtwo$};
	\draw (f) -- (X);
	\draw (f) -- (g1);
	\draw (f) -- (Y2);
	\draw (g1) -- (Z1);
	\draw (g1) -- (Z2);
\end{tikzpicture} } &
\scalebox{0.85}{ \begin{tikzpicture}
	\node[draw,circle] (f) at (0,0) {$\opOp$};
	\node[draw,circle,inner sep=0pt,minimum size=5ex] (g1) at (.75,-1) {$\optwo$};
	\node (X) at (0,1) {$\System$};
	\node (Z3) at (-.25,-2) {$\Subtwoone$};
	\node (Z2) at (.75,-2) {$\Subtwotwo$};
	\node (Z1) at (1.75,-2) {$\Subtwothree$};
	\node (Y2) at (-1.25,-2) {$\Subone$};
	\draw (f) -- (X);
	\draw (f) -- (g1);
	\draw (f) -- (Y2);
	\draw (g1) -- (Z1);
	\draw (g1) -- (Z2);
	\draw (g1) -- (Z3);
\end{tikzpicture} }&
\scalebox{0.85}{ \begin{tikzpicture}
	\node[draw,circle] (f) at (0,0) {$\opOp$};
	\node[draw,circle,inner sep=0pt,minimum size=5ex] (g1) at (-1.25,-1) {$\opone$};
	\node (X) at (0,1) {$\System$};
	\node (Z1) at (-2,-2) {$\Suboneone$};
	\node (Z2) at (-.75,-2) {$\Subonetwo$};
	\draw (f) -- (X);
	\draw (f) -- (g1);
	\draw (g1) -- (Z1);
	\draw (g1) -- (Z2);
	\node[draw,circle,inner sep=0pt,minimum size=5ex] (g1) at (1.25,-1) {$\optwo$};
	\node (Z3) at (.25,-2) {$\Subtwoone$};
	\node (Z4) at (1.25,-2) {$\Subtwotwo$};
	\node (Z5) at (2.25,-2) {$\Subtwothree$};
	\draw (f) -- (X);
	\draw (f) -- (g1);
	\draw (g1) -- (Z3);
	\draw (g1) -- (Z4);
	\draw (g1) -- (Z5);
\end{tikzpicture}}
\end{array}
\label{eq:trees}
\end{equation}
Together with the original operation, these represent four views of the same system at different levels of granularity; compare, e.g., \cite[Fig.\ 2]{Leveson}. This reveals an important point: an operad provides a collection of interrelated models that fit together to represent a complex system.

The relationship between models is constrained by the \emph{principle of compositionality}: the whole is determined by its parts \emph{and} their organization. Here, the whole is the root, the parts are the leaves, and each tree is an organizational structure. Formally, \emph{associativity axioms}, which generalize those of ordinary categories, enforce compositionality. For example, composing the left-hand tree above with $\optwo$ must give the same result as composing the center tree with $\opone$.  Both give the tree on the right, since they are built up from the same operations. 
In day-to-day modeling these axioms are mostly invisible, ensuring that everything ``just works'', but the formal definitions \cite[11.2]{Yau} provide explicit requirements and desiderata for modeling languages ``under the hood''.

Operads encourage principled approaches to emergence by emphasizing the organization of a system. Colloquially speaking, an emergent system is ``more than the sum of its parts''; operations provide a means to describe these nonlinearities. This does not explain emergent phenomena, which requires detailed semantic modeling, but begins to break up the problem with separate (but related) representation of components and their interactions. The interplay between these elements can be complex and unexpected, even when the individual elements are quite simple.\footnote{For example, diffusion rates (components) and activation/inhibition (interactions) generate zebra's stripes in Turing's model of morphogenesis \cite{TuringMorph}.} 
Compositional models may develop and exhibit emergence as interactions between components are organized, in much the same way as the systems they represent.

\subsection{The API view}
\label{subsec:API_view}
For most applications, trees bear labels: fault trees, decision trees, syntax trees, dependency trees and file directories, to name a few. A tree's labels indicate its semantics either explicitly with numbers and symbols or implicitly through naming and intention.

In an operad, nodes identify operations while edges---called \emph{types}---restrict the space of valid compositions. This is in analogy to type checking in strongly-typed programming languages, where we can only compose operations when types match. 
In the API view, the operations are abstract method declarations:\vspace*{-7pt}
\begin{quote}\begin{tabular}{l}
	\texttt{def op(x1 : Sub1, x2 : Sub2) : System},\\
	\texttt{def op1(y1 : Sub11, y2 : Sub12) : Sub1},\\
	\texttt{def op2(z1 : Sub21, z2 : Sub22, z3 : Sub23) : Sub2}.\\
\end{tabular}\end{quote}\vspace*{-5pt}
Composites are essentially scripted methods defined in the API. For example, 
\vspace*{-7pt}
\begin{quote}
	\texttt{def treeLeft(y1 : Sub11, y2 : Sub12, x2 : Sub2) : System\\ \hspace*{6mm} = op(op1(y1, y2), x2)},
\end{quote}\vspace*{-5pt}
is a script for left-most tree above. However, the compiler will complain with an invalid syntax error for any script where the types don't match, say \vspace*{-7pt}
\begin{quote}
	\texttt{def badTree(y1 : Sub11, y2 : Sub12, x2 : Sub2) : System\\ \hspace*{6mm} = op(x2 ,op1(y1, y2))},
\end{quote} \vspace*{-5pt}

If an operad is an API---a collection of abstract types and methods---then an \emph{operad algebra} $\Alg$ is a concrete implementation. 
An algebra declares: 1) a set of {\it instances} for each type; 2) a {\it function} for each operation, taking instances as arguments and returning a single instance for the composite system. That is, $\Alg \maps \OO \to \Set$ has: \vspace*{-7pt}
    \begin{itemize}
    \item for each type $\X \in T$, a set $\Alg(\X)$ of \define{instances} of type $\X$, and
    \item for each operation $\fop \maps \< \Xitype\> \to \Y$, the function $\Alg(\fop)$ \define{acts} 
    on input elements $ a_i \in \Alg(\Xitype)$
    to obtain a single output element
    $\Alg(\fop)(a_1, \dots, a_n) \in \Alg(\Y).$
\end{itemize} \vspace*{-5pt}
Required coherence rules \cite[13.2]{Yau} are analogous to the definition of a functor into  $\Set$ \cite[I.3]{MacLane}.
For example, we might declare a state space for each subsystem, and a function to calculate the overall system state given subsystem states. Alternatively, we might assign key performance indicators (KPIs) for each level in a system and explicit formulae to aggregate them. The main thing to remember is: just as an abstract method has many implementations, an operad has many algebras. Just like an API, the operad provides a common syntax for a range of specific models, suited for specific purposes.

Unlike a traditional API, an operad provides an explicit framework to express and reason about semantic relationships between \emph{different} implementations.
These different implementations are linked by type-indexed mappings between instances called 
\emph{algebra homomorphisms}. For example, we might like to extract KPIs from system state. The principle of compositionality places strong conditions on this extraction: the KPIs extracted from the overall system state must agree with the KPIs obtained by aggregating subsystem KPIs. That is, in terms of trees
and in pseudocode: \vspace*{-7pt}
{
\begin{center}
\begin{tabular}{p{5.5cm}cp{5cm}}
    \begin{tabular}{c}
    \scalebox{0.9}{
    	\begin{tikzpicture}
    		\node[draw] (f) at (0,0) {$\KPI(\opOp)$\strut};
    		\node[draw,rounded corners] (g1) at (-1,-1.5) {$\extr(\Subone)$\strut};
    		\node[draw,rounded corners] (g2) at (1,-1.5) {$\extr(\Subtwo)$\strut};
    		\node (X) at (0,1) {$\KPI(\System)$};
    		\node (Z1) at (-1,-2.5) {$\State(\Subone)$};
    		\node (Z2) at (1,-2.5) {$\State(\Subtwo)$};
    		\draw (f) -- (X);
    		\draw (f) -- node[left] {$\KPI(\Subone)$} (g1);
    		\draw (f) -- node[right] {$\KPI(\Subtwo)$} (g2);
    		\draw (g1) -- (Z1);
    		\draw (g2) -- (Z2);
    	\end{tikzpicture} }
	\end{tabular}
 & = &
	\begin{tabular}{c}
	\scalebox{0.85}{\begin{tikzpicture}
		\node[draw] (f) at (0,-1.5) {$\State(\opOp)$\strut};
		\node[draw,rounded corners] (g1) at (0,0) {$\extr(\System)$ \strut};
		\node (X) at (0,1) {$\KPI(\System)$};
		\node (Z1) at (-1,-2.5) {$\State(\Subone)$};
		\node (Z2) at (1,-2.5) {$\State(\Subtwo)$};
		\draw (g1) -- (X);
		\draw (g1) -- node[right] {$\State(\System)$} (f);
		\draw (f) -- (Z1);
		\draw (f) -- (Z2);
	\end{tikzpicture}}
	\end{tabular}
	\\ 
	\hfill\texttt{KPI(op)(extr(x1), extr(x2)) } & \texttt{==} & \texttt{extr(State(op)(x1, x2))}.\\
\end{tabular}
\end{center}
}\vspace*{-5pt}
\noindent   For any state instances for $\Subone$ and $\Subtwo$ at the base of the tree, the two computations must produce the same KPIs for the overall system at the top of the tree. Here $\KPI(\opOp)$ and $\State(\opOp)$ implement $\opOp$ in the two algebras, while $\extr(-)$ are \emph{components} of the algebra homomorphism to extract KPIs.
Similar to associativity, these compositionality conditions guarantee that extracting KPIs ``just works'' when decomposing a system hierarchically. 
  
\subsection{The equational view} 
\label{subsec:equate_view}
We have just seen an equation between trees that represent implementations. Because an operad can be studied without reference to an implementation, we can also define equations between abstract trees. This observation leads to another view of an operad: as a system of equations.

The first thing to note is that equations occur within the sets of operations $\OO(\Xone,\ldots,\Xn;\mathtt{Y})$; an equation between  two operations only makes sense if the input and output types match. Second, if one side of an equation $\fop=\fop'$ occurs as a subtree in a larger operation $\mathtt{g}$, substitution generates a new equation $\mathtt{g}=\mathtt{g'}$. Two trees are equal if and only if they are connected by a chain of such substitutions (and associativity equations). 
In general, deciding whether two trees are equal (the word problem) may be intractable.
Third, we can often interpret composition of operations as a normal-form computation:
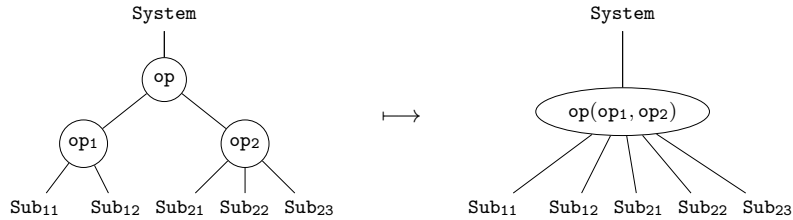
\[\begin{array}{ccccc}
	\begin{array}{c}
	\scalebox{0.85}{
		\begin{tikzpicture}
			\node[draw,circle] (f) at (0,0) {$\opOp$};
			\node[draw,circle,inner sep=0pt,minimum size=5ex] (g1) at (-1.25,-1) {$\opone$};
			\node (X) at (0,1) {$\System$};
			\node (Z1) at (-2,-2) {$\Suboneone$};
			\node (Z2) at (-.75,-2) {$\Subonetwo$};
			\draw (f) -- (X);
			\draw (f) -- (g1);
			\draw (g1) -- (Z1);
			\draw (g1) -- (Z2);
			\node[draw,circle,inner sep=0pt,minimum size=5ex] (g1) at (1.25,-1) {$\optwo$};
			\node (Z3) at (.25,-2) {$\Subtwoone$};
			\node (Z4) at (1.25,-2) {$\Subtwotwo$};
			\node (Z5) at (2.25,-2) {$\Subtwothree$};
			\draw (f) -- (X);
			\draw (f) -- (g1);
			\draw (g1) -- (Z3);
			\draw (g1) -- (Z4);
			\draw (g1) -- (Z5);
		\end{tikzpicture}}
	\end{array}
	& \longmapsto & 	
	\begin{array}{c} 
		\scalebox{0.85}{\begin{tikzpicture}
			\node[draw,ellipse] (f) at (0,-.5) {$\opOp(\opone, \optwo)$};
			\node (X) at (0,1) {$\System$};
			\node (Z1) at (-2,-2) {$\Suboneone$};
			\node (Z2) at (-.75,-2) {$\Subonetwo$};
			\draw (f) -- (X);
			\draw (f) -- (Z1);
			\draw (f) -- (Z2);
			\node (Z3) at (.25,-2) {$\Subtwoone$};
			\node (Z4) at (1.25,-2) {$\Subtwotwo$};
			\node (Z5) at (2.25,-2) {$\Subtwothree$};
			\draw (f) -- (X);
			\draw (f) -- (Z3);
			\draw (f) -- (Z4);
			\draw (f) -- (Z5);
		\end{tikzpicture}}
	\end{array}
\end{array}\]
We then compare composed operations directly to decide equality.  For example, there is an operad whose operations are matrices. Composition computes a normal form for a composite operation by block diagonals and matrix multiplication,
\[\begin{array}{ccccc}
	\begin{array}{l} 
		op:n\times (m_1+m_2)\\
		op_1:m_1\times (k_{11}+k_{12})\\
		op_2:m_2\times (k_{21}+k_{22}+k_{23})\\
	\end{array}
	& \longmapsto & 	
	\Big(\ op\ \Big)\cdot 
	\begin{pmatrix} \ op_1\  & 0 \\\\ 0 & \ op_2\ \\\end{pmatrix}.
\end{array}\]
Operad axioms constrain composition. For example,  the axiom mentioned in \ref{sec:operads}\ref{subsec:tree_view} corresponds to: \[
 	\Big(\ op\ \Big)
 	\cdot
 	\begin{pmatrix} 
 		op_1 & 0 \\\\
 		0 & I_{m_2} \\
 	\end{pmatrix}
 	\cdot
 	\begin{pmatrix} 
 		I_{k_{11}} & 0 & 0\\
 		0 & I_{k_{12}} & 0 \\
 		0 & 0 & op_2\\
 	\end{pmatrix}
 	=
	\Big(\ op\ \Big)
 	\cdot
 	\begin{pmatrix} 
 		I_{m_1} & 0 \\\\
 		0 & op_2 \\
 	\end{pmatrix}
 	\cdot
 	\begin{pmatrix} 
 		op_1 & 0 & 0 & 0\\
 		0 & I_{k_{21}} & 0 & 0\\
 		0 & 0 & I_{k_{22}} & 0 \\
 		0 & 0 & 0 & I_{k_{23}} \\
 	\end{pmatrix}.
 \]
 \newcommand{\bigexists}{\raisebox{-1ex}{\scalebox{2}{$\exists$}}}

The key point is that any algebra that implements the operad must satisfy \emph{all} of the equations that it specifies.    
Type discipline controls which operations can compose; equations between operations control the resulting composites.
Declaring equations between operations provides additional contracts for the API.  For instance, any unary operation $\fop \maps \X \to \X$ (a loop) generates an infinite sequence of composites $\id_\X,\fop,\fop^2,\fop^3,\ldots$. Sometimes this is a feature of the problem at hand, but in other cases we can short-circuit the infinite regress with assumptions like idempotence ($\fop^2=\fop$) or cyclicity ($\fop^n=\id_\X$) and ensure that algebras contain no infinite loops.

\subsection{The systems view} 
\label{subsec:mapmaker}

When we apply operads to study systems, we often think of an operation $\fop \maps \langle\Xitype\rangle\to Y$ as a system architecture. Intuitively $\mathtt{Y}$ is the system and the $\Xone,\ldots,\Xn$ are the components, but this is a bit misleading. It is better to think of types as boundaries or interfaces, rather than systems. Instead, $\fop$ is the system architecture, with component interfaces $\Xitype$ and environmental interface $\mathtt{Y}$. Composition formalizes the familiar idea \cite[Fig.\ 2]{Leveson} that one engineer's system is the next's component; it acts by nesting subsystem architectures within higher-level architectures. 

Once we establish a system architecture, we would like to use this structure to organize our data and analyses of the system. Moreover, according to the principle of compositionality, we should be able to construct a system-level analysis from an operation by analyzing the component-level inputs and synthesizing these descriptions according to the given operations. 

The process of extracting computations from operations is called \emph{functorial semantics}, in which a model is represented as a mapping $\mathsf{M}\maps\mathbf{Syntax}\longrightarrow\mathbf{Semantics}$. 
The syntax defines a system-specific architectural design. Semantics are universal and provide a computational context to interpret specific models. Matrices, probabilities, proofs, and dynamical equations all have their own rules for composition, corresponding to different semantic operads.

The mapping $\mathsf{M}$ encodes, for each operation, the data, assumptions and design artifacts (e.g., geometric models) needed to construct the relevant computational representations for the architecture, its components and the environment. From this, the system model as a whole is determined by composition in the semantic context.
The algebras ($\State$,$\KPI$) described in \ref{sec:operads}\ref{subsec:API_view}  are typical examples, with syntax $\OO$ and taking semantic values in sets and functions. The mappings themselves, called \emph{functors}, map types and operations ($\System$, $\opOp$) to their semantic values, while preserving how composition builds up complex operations.

The functorial perspective allows complementary models--e.g.\ system state vs.\ KPIs--to be attached to the same design. This includes varying the semantic context as well as modeling details; see Sec.\ \ref{sec:anal} for examples of non-deterministic semantics. Though functorial models may be radically different, they describe the \emph{same system}, as reflected by the overlapping syntax.

In many cases, relevant  models are \emph{not} independent, like system state and KPIs. Natural transformations, like the extraction homomorphism in \ref{sec:operads}\ref{subsec:API_view},   provide a means to align ostensibly independent representations. 
Since models are mappings, we often visualize natural transformations as a two-dimensional cells:
	\begin{equation}
	\xymatrix{\OO \ar@/^2.5ex/[rr]^{\State} \ar@/_2.5ex/[rr]_{\KPI} \ar@{}[rr]|-{\rotatebox{270}{\Large $\Rightarrow$}}&& \Set .}
	\label{eq:algHom}
	\end{equation}
Formal conditions guarantee that when moving from syntax to semantics \cite[13.2]{Yau} or between representations \cite[2.3.5]{Leinster}, reasoning about how systems decompose hierarchically ``just works.''

Since functors and higher cells assure coherence with hierarchical decomposition, we can use them to build up a desired model in stages, working backwards from simpler models:
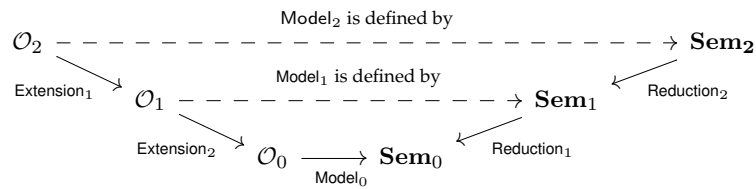
\[\xymatrix@R=1ex{
\mathcal{O}_2 \ar[rd]_{\textsf{Extension}_1\ \ } \ar@{-->}[rrrrr]^-{\mathsf{Model}_2\textrm{ is defined by}} &&&&& \mathbf{Sem_2} \ar[dl]^{\hspace{3ex}\textsf{Reduction}_2}\\
& \mathcal{O}_1 \ar@{-->}[rrr]^{\textsf{Model}_1\textrm{ is defined by}} \ar[rd]_{\textsf{Extension}_2\ \ }&&& \mathbf{Sem}_1
\ar[dl]^{\hspace{3ex}\textsf{Reduction}_1}&\\
&& \mathcal{O}_0 \ar[r]_{\textsf{Model}_0} & \mathbf{Sem}_0 && \\
}\]
This is a powerful technique for at least two reasons. First, complexity can be built up in stages by layering on details. Second, complex models built at later stages are partially validated through their coherence with simpler ones. The latter point is the foundation for lazy evaluation: 
many coarse models can be explored before ever constructing expensive models. 

Separating out the different roles within a model encourages efficiency and reuse. An architecture (operation) developed for one analysis can be repurposed with strong coherence between models (algebra instances) indexed by the same conceptual types. 
The syntax/semantics distinction also helps address some thornier meta-modeling issues. For example, syntactic types can distinguish conceptually distinct entities while still mapping to the same semantic entities. We obtain the flexibility of structural or duck typing in the semantics without sacrificing the type safety provided by the syntax. 

 \section{Main examples}
 \label{sec:examples} 
 
 Though operads are general tools \cite{Leinster, MSS, Yau}, we focus on two classes of operads applied to system design: wiring diagram operads and network operads. These are complementary. Wiring diagrams provide a top-down view of the system, whereas network operads are bottom-up.  This section introduces 3 examples that help ground the exposition as in Fig.\ \ref{fig:org}.
 
 \subsection{Specification}
\label{subsec:exSailboat}

Network operads describe atomic types of systems and ways to link them together with operations.
These features enable: (1) specification of atomic building blocks for a domain problem; and (2) bottom up synthesis of designs from atomic systems and links.
 A general theory of network operads \cite{NetworkModels,  NMPetri, Moeller, MonGroth} was recently developed under the Defense Advanced Research Projects Agency (DARPA)  Complex Adaptive System Composition and Design Environment (CASCADE) program. 
Minimal data can be used to specify a functor--called a network model \cite[ 4.2]{NetworkModels}--which constructs a network operad \cite[ 7.2]{NetworkModels} customized to a domain problem.
 \begin{figure}[!ht]
     \centering
     \begin{subfigure}[b]{0.5\textwidth}
         \centering
          \begin{tabular}{| c | c | c | c | c | c | c | }
  \hline 
  & $\BOAT$ &  $\HELO$ & $\FWUAV$ & $\QUAD$  \\
  \hline
  $\CUTTER$ & 1 &  1 & 1 &  1 \\
  \hline
   $\BOAT$ &  &  & 1  & 1 \\
  \hline
  $\FWS$ &  & &  & 1\\
  \hline
  $\FWSAR$ & &  &   & 1\\
  \hline
  $\HELO$ & &  &  &  1 \\
  \hline
  \end{tabular}
  \caption{Examples of carrying relationships in $\OO_{Sail}$}
  \label{tab:synport}
  \vspace*{-4pt}
     \end{subfigure}
     \hfill
     \begin{subfigure}[b]{0.48\textwidth}
         \centering
         {
\begin{tikzpicture}
	\begin{pgfonlayer}{nodelayer}
		\node [style=dot, draw=ForestGreen, fill=ForestGreen] (1) at (2, 1) {}; 
		\node [style=dot,  draw=red, fill=red] (2) at (2.4, 1) {}; 
		\node [style=dot, draw=blue, fill=blue] (3) at (2, 0.6) {}; 
		\node [style=dot, draw=blue, fill=blue] (4) at (2.4, 0.6) {};
		\node [style=none] (a) at (1.4,1.2) {};
		\node [style=none] (b) at (3.0,1.2) {};
		\node [style=none] (c) at (2.2,-.2) {};
		\node [style=dot, draw=ForestGreen,fill=ForestGreen] (D1) at (1.7, 1.6) {};
		\node [style=dot, draw=blue, fill=blue] (D2) at (1.9, 1.6) {};
		\node [style=none] (D) at (1.8, 1.15) {};
		\node [style=none] (D') at (1.8, 1.5) {};
		\node [style=dot, draw=red, fill=red] (E1) at (2.5, 1.6) {};
		\node [style=dot, draw=blue, fill=blue] (E2) at (2.7, 1.6) {};
		\node [style=none] (E) at (2.6, 1.15) {};
		\node [style=none] (E') at (2.6, 1.5) {};
		\node [style=none] (c') at (2.2,-.15) {};
		\node [style=none] (C) at (2.2, -0.5) {};
		\node [style=dot, draw=ForestGreen, fill=ForestGreen] (C1) at (2.1, -0.6) {};
		\node [style=dot, draw=red, fill=red] (C2) at (2.3, -0.6) {};
		\node [style=dot, draw=blue, fill=blue]  (C3) at (2.1, -0.8) {};
		\node [style=dot, draw=blue, fill=blue] (C4) at (2.3, -0.8) {};
		\node [style=none] (Cut) at (3.92, 0.8) {{\textcolor{ForestGreen}{$\bullet$}} $\CUTTER$};
		\node [style=none] (Helo) at (4, 0.4) {{\textcolor{red}{$\bullet$}} $\HELO$};
		\node [style=none] (QD) at (3.845, -0.02) {{\textcolor{blue}{$\bullet$}} $\QUAD$};
		
	\end{pgfonlayer}
	\begin{pgfonlayer}{edgelayer}
		\draw [->, semithick] (2) to (1);
		\draw [->, semithick] (3) to (1);
		\draw [->, semithick] (4) to (2);
		\draw [style=simple]  (a) to (b);
		\draw [style=simple]  (a) to (c);
		\draw [style=simple]  (b) to (c);
		\draw [style=simple] (D) to (D');
		\draw [style=simple] (E) to (E');
		\draw [style=simple] (c') to (C);
	\end{pgfonlayer}
\end{tikzpicture}
}
\caption{Operation $\fop \in {\OO}_{Sail}$ to specify carrying}
  \label{fig:exSailboat}
  \vspace*{-4pt}
     \end{subfigure}
         \caption{
Which types are allowed to carry other types--indicated with $1$--specify an operad $\OO_{Sail}$; $\fop$ specifies that a  $\HELO$ ({\textcolor{red}{$\bullet$}}) and a $\QUAD$ ({\textcolor{blue}{$\bullet$}}) are carried by a $\CUTTER$ ({\textcolor{ForestGreen}{$\bullet$}}) and another $\QUAD$ ({\textcolor{blue}{$\bullet$}}) is carried on the $\HELO$ ({\textcolor{red}{$\bullet$}}).}
\label{fig:sailnetop}
\vspace*{-5pt}
\end{figure}

The first example illustrates designs of search and rescue (SAR) architectures.
The domain problem was inspired by the 1979 Fastnet Race and the 1998 Sydney to Hobart Yacht Race and we refer to it as the sailboat problem.    It illustrates how network operads facilitate the specification of a model with combinatorial data called a network template.  For example, Fig.\ \ref{fig:sailnetop} shows the carrying relationships between different system types to model (e.g., a $\BOAT$ can carry a $\FWUAV$ (Unmanned Aerial Vehicle) but a $\HELO$ cannot). This data specifies a network operad $\OO_{Sail}$ whose: (1) objects are lists of atomic system types; (2) operations describe systems carrying other systems; and (3) composition combines carrying instructions.   We discuss this example in greater detail in Sec.\ \ref{sec:cookbook}.

\begin{figure}[t]
	\begin{center}
		\begin{tabular}{|c|c|}
			\hline
			\textbf{Functional Decomposition} & \textbf{Control Decomposition}\\
			\scalebox{.65}{
				\begin{tikzpicture}
				\node (1) at (3,0) {$\Lab$};
				\node (2) at (2,-.5) {$\BoxTemp$};
				\node (3) at (1,0) {$\Bath$};
				\node (4) at (-1,0) {$\Chassis$};
				\node (5) at (-2,-.5) {$\Optics$};
				\node (6) at (-3,0) {$\Interfer$};
				\node[blue] (11) at (2,1.5) {$\TempSys$};
				\node[blue] (12) at (-2,1.5) {$\LengthSys$};
				\node (0) at (0,3) {$\LSIob$};
				\draw (1) to (11);
				\draw (2) to (11);
				\draw (3) to (11);
				\draw (4) to (12);
				\draw (5) to (12);
				\draw (6) to (12);
				\draw (11) to (0);
				\draw (12) to (0);
				\node (f) at (-3.25,2.25) {$\varphiOp \left\{\rule{0cm}{1.155cm}\right.$};
				\node (ell) at (-4,.5) {$\lambdaOp \left\{\rule{0cm}{1.25cm}\right.$};
				\node (ell) at (4.2,.5) {$\left.\rule{0cm}{1.25cm}\right\}\tauOp$};
				\end{tikzpicture}
			}
			&
			\scalebox{.65}{
				\begin{tikzpicture}
				\node (1) at (-3.75,0.25) {$\Lab$};
				\node (2) at (-2.75,-.5) {$\BoxTemp$};
				\node (3) at (-1.25,-.5) {$\Optics$};
				\node (4) at (-.25,0.25) {$\Interfer$};
				\node (5) at (1,-.5) {$\Chassis$};
				\node (6) at (3,-.5) {$\Bath$};
				\node[red] (11) at (-2,1.5) {$\Sensors$};
				\node[red] (12) at (2,1.5) {$\Actuators$};
				\node (0) at (0,3) {$\LSIob$};
				\draw (1) to (11);
				\draw (2) to (11);
				\draw (3) to (11);
				\draw (4) to (11);
				\draw (5) to (12);
				\draw (6) to (12);
				\draw (11) to (0);
				\draw (12) to (0);
				\node (c) at (-3,2.25) {$\kappaOp \left\{\rule{0cm}{1.15cm}\right.$};
				\node (s) at (-5,.5) {$\sigmaOp \left\{\rule{0cm}{1.25cm}\right.$};
				\node (a) at (4,.5) {$\left.\rule{0cm}{1.25cm}\right\}\alphaOp$};
				\end{tikzpicture}
			}\\\hline
			\multicolumn{2}{|c|}{\textbf{Operad Equation:} $\varphiOp(\lambdaOp,\tauOp)=\kappaOp(\sigmaOp,\alphaOp)$}\\
			\multicolumn{2}{|c|}{
				\scalebox{.65}{
					\begin{tikzpicture}
					\node[draw,rectangle,rounded corners=5pt] (optics) at (-3,1) {$\Optics$};
					\node[draw,rectangle,rounded corners=5pt] (chassis) at (0,2) {$\Chassis$};
					\node[draw,rectangle,rounded corners=5pt] (laser) at (-3,0) {$\Interfer$};
					\node[draw,rectangle,rounded corners=5pt] (roomtemp) at (3,-1.5) {$\Lab$};
					\node[draw,rectangle,rounded corners=5pt] (boxtemp) at (0,-1.5) {$\BoxTemp$};
					\node[draw,rectangle,rounded corners=5pt] (bath) at (3,1) {$\Bath$};
					\node[draw,rectangle,rounded corners=5pt,densely dotted] (target0) at (0,0) {\rule{0cm}{7cm} \rule{10cm}{0cm}};
					\node (target0label) at (-4.9,3.8) {$\LSIob$}; 
					\node[draw,circle] (laserdot) at (-1.5,0) {};
					\draw (laser) to (laserdot);
					\draw[rounded corners] (chassis.-45) to (chassis.-45|-optics) to node[above, near end,rotate=40] {\small $\laser$} (chassis.west|-laserdot) to (laserdot);
					\draw[rounded corners] (boxtemp.180) to (boxtemp-|laserdot) to (laserdot);
					\node[draw,circle] (intensdot) at (-1.75,2) {};
					\node (intensity) at (-1.75,4.2) {\small $\intens$};
					\draw[dashed,rounded corners] (intensdot) to node[above,very near end] {\small $\intens$} (chassis-|optics) to (optics);
					\draw[dashed,rounded corners] (intensdot) to (intensity);
					\draw[dashed,rounded corners] (intensdot) to (chassis);
					\draw[rounded corners] (chassis.-135) to (chassis.-135|-optics) to node[below] {\small $\focus$} (optics);
					\node (position) at (0,4.2) {\small $\drive$};
					\draw[dashed] (position) to (chassis);
					\node (fringes) at (-6.25,0) {\small $\fringe$};
					\draw[dashed] (laser) to (fringes);
					\draw[rounded corners] (boxtemp) to node[above] {\small $\heat_1$} (roomtemp);
					\draw[rounded corners] (boxtemp) to (0,-.5) to (3,-.5) to node[right] {\small$\heat_2$} (bath);
					\node (set) at (6.25,1) {\small $\setPt$};
					\draw[dashed] (set) to (bath);
					\node (h20) at (3,4.2) {\small $\water$};
					\draw (h20) to (bath);
					\node (tempSignal) at (1.5,-4.25) {\small $\temp$};
					\draw[dashed, rounded corners] (roomtemp) to (3,-2) to (1.6,-2) to (1.6,-4);
					\draw[dashed, rounded corners] (boxtemp) to (0,-2) to (1.4,-2) to (1.4,-4);
					\node[blue] (target1label) at (3.8,-3) {$\TempSys$}; 
					\draw[densely dotted,thick,rounded corners,blue] (4.6,0) to (4.6,1.5) to (1.75,1.5) to (-1.25,-1.25) to (-1.25,-2.8) to (4.6,-2.8) to (4.6,0);
					\node[blue] (target2label) at (-3,3) {$\LengthSys$}; 
					\draw[densely dotted,thick,rounded corners,blue] (-4,0) to (-4,2.8) to (1,2.8) to (1,1.3) to (-1.5,-1) to (-4,-1) to (-4,0);
					\draw[densely dotted,thick,rounded corners,red] (-4.4,0) to (-4.4,1.6) to (-2,1.6) to (4.3,-1) to (4.3,-2.5) to (-4.4,-2.5) to (-4.4,0);
					\node[red] (target1label) at (-3.8,-2.7) {$\Sensors$}; 
					\draw[thick,densely dotted,rounded corners,red] (2,2.5) to (4.3,2.5) to (4.3,.5) to (1.7,.5) to (-1.3,1.75) to (-1.3,2.5) to (2,2.5);
					\node[red] (target2label) at (4,2.7) {$\Actuators$}; 
					\end{tikzpicture}
				}
			}\\ \hline
		\end{tabular}
		\vspace*{-15pt}
	\end{center}
	\caption{An equation in a wiring diagram operad operations expresses a common refinement of 
	hierarchies.}
	\label{fig:decomps}
	\vspace*{-15pt}
\end{figure}
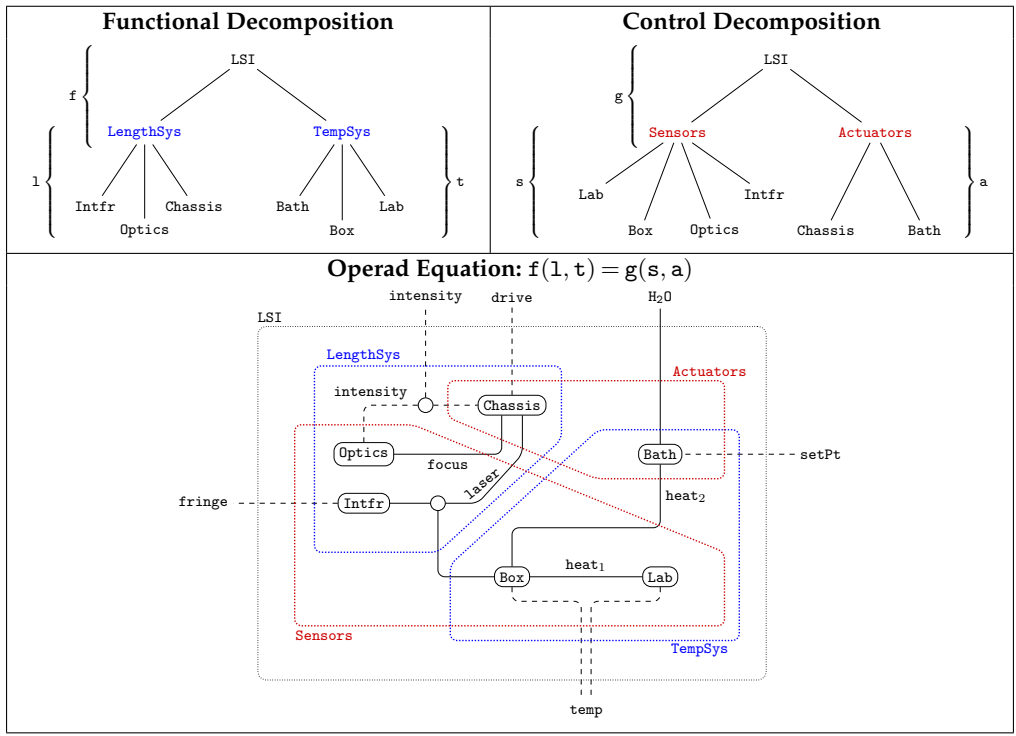

\subsection{Analysis}
\label{subsec:wireOPIntro}

A wiring diagram operad describes the interface each system exposes, making it clear what can be put together
\cite{OpWire, SpivakTan, OpWireYau}.
The designer has to specify precisely how information and physical quantities are shared among components, while respecting their interfaces.
The operad facilitates top-down analysis of a design by capturing different ways to decompose a composite system. 

The second example analyzes a precision-measurement system called the Length Scale Interferometer (LSI) with wiring diagrams.  
It helps illustrate the qualitative features of operads over and above other modeling approaches and the potential to exploit their analytic power to separate concerns.  Figure \ref{fig:decomps} illustrates joint analysis of the LSI to address different aspects of the design problem: functional roles of  subsystems and control of the composite system. This analysis example supports these illustrations in Sec.\ \ref{sec:anal}.
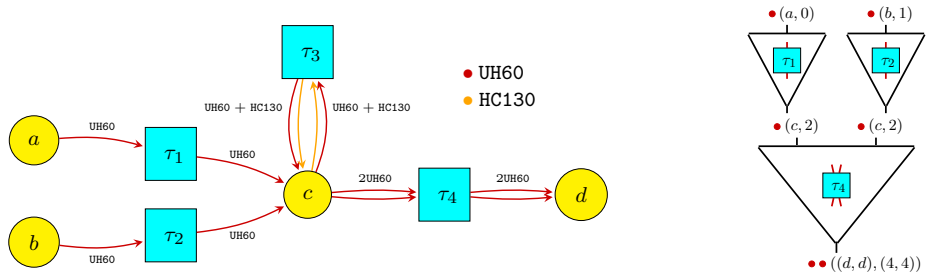
\begin{figure}[!ht]
     \centering
     \begin{subfigure}[b]{0.62\textwidth}
         \centering
   \scalebox{0.9}{
\begin{tikzpicture}
	\begin{pgfonlayer}{nodelayer}
		\node [style=species] (A) at (-5, 0.8) {$\;\;a\;\;$};
		\node [style=species] (B) at (-5, -0.7) {$\;\;b\;\;$};
		\node [style=species] (C) at (-1, 0) {$\;\;c\;\;$};
		\node [style=species] (D) at (3, 0) {$\;\;d\;\;$};
		\node [style=transition] (tau1) at (-3, 0.6) {$\;\phantom{\Big{|}}\tau_1\;$};
		\node [style=transition] (tau2) at (-3, -0.6) {$\;\phantom{\Big{|}}\tau_2\;$};
        \node [style=transition] (tau3) at (-1, 2.1) {$\;\phantom{\Big{|}}\tau_3\;$};
        \node [style=transition] (tau4) at (1, 0) {$\;\phantom{\Big{|}}\tau_4\;$};
		\node [style=none] (HH) at (1.715, 1.8) {\small {\textcolor{red}{$\bullet$}} $\HH$};
		\node [style=none] (HH) at (1.8, 1.4) {\small {\textcolor{Orange}{$\bullet$}} $\HC$};
	\end{pgfonlayer}
	\begin{pgfonlayer}{edgelayer}
		\draw [style=inarrow, draw=red] (A) to [bend left=10] node[midway, above] {\tiny$\HH$} (tau1) ;
 		\draw [style=inarrow, draw=red] (tau1) to [bend left=10]  node[midway, above] {\tiny$\HH$} 
 		(C);
	    \draw [style=inarrow, draw=red] (B) to [bend right=10]  node[midway, below]{\tiny$\HH$  }  (tau2);
        \draw [style=inarrow, draw=red] (tau2) to [bend right=10]  node[midway, below]{\tiny$\HH$} (C); 
	    \draw [style=inarrow, draw=red] (C) to [bend right=20]  node[pos=0.7,right
	    ]{\tiny$\HH +\HC$} (tau3);
	    \draw [style=inarrow, draw=Orange] (C) to [bend right=10] (tau3);
        \draw [style=inarrow, draw=red] (tau3) to [bend right=20]  node[pos=0.3,left
	    ]{\tiny$\HH +\HC$} (C);
	    \draw [style=inarrow, draw=Orange] (tau3) to [bend right=10] (C);
	    \draw [style=inarrow, draw=red] (C)  to [bend left=5] node[midway, above]{\tiny$2\HH$} (tau4);
	    \draw [style=inarrow, draw=red] (C) to [bend right=5]  (tau4);
        \draw [style=inarrow, draw=red] (tau4) to [bend left=5] node[midway, above]{\tiny$2\HH$} (D);
        \draw [style=inarrow, draw=red] (tau4) to [bend right=5]  (D);
	\end{pgfonlayer}
\end{tikzpicture}
}      
  \caption{Specification of primitive tasks $:=$ transitions}
  \label{fig:PetriInputs}
  \vspace*{-4pt}
     \end{subfigure}
     \hfill
     \begin{subfigure}[b]{0.35\textwidth}
         \centering
  \scalebox{0.65}{
\begin{tikzpicture}
	\begin{pgfonlayer}{nodelayer}
		\node [style=none] (a) at (0.4,3) {};
		\node [style=none] (b) at (2.0,3) {};
		\node [style=none] (c) at (1.2,1.5) {};
		\node [style=none] (Atau1) at (1.2,2.85) {};
		\node [style=minitransition] (tau1) at (1.2,2.45) {\large{$\phantom{{|}}\tau_1$}};
		\node [style=none] (Ctau1) at (1.2,2.05) {};
		\node [style=none] (d) at (2.4,3) {};
		\node [style=none] (e) at (4.0,3) {};
		\node [style=none] (f) at (3.2,1.5) {};
		\node [style=none] (Btau2) at (3.2,2.85) {};
		\node [style=minitransition] (tau2) at (3.2,2.45) {\large{$\phantom{{|}}\tau_2$}};
		\node [style=none] (Ctau2) at (3.2,2.05) {};
		\node [style=none] (h) at (0.6,0.7) {};
		\node [style=none] (j) at (3.8,0.7) {};
		\node [style=none] (i) at (2.2,-1.3) {};
		\node [style=none] (C1tau4) at (2.1,0.35) {};
		\node [style=none] (C2tau4) at (2.3,0.35) {};
		\node [style=minitransition] (tau4) at (2.2,-0.1) {\large{$\phantom{{|}}\tau_4$}};
		\node [style=none] (D1tau4) at (2.1,-0.55) {};
		\node [style=none] (D2tau4) at (2.3,-0.55) {};
		\node [style=dot,  fill=red,  draw=red] (D1) at (0.9, 3.4) {};
		\node [style=none] (D2) at (1.4, 3.4) {\small{$(a, 0)$}};
		\node [style=none] (D') at (1.2, 3.2) {};
		\node [style=none] (D) at (1.2, 2.98) {};
		\node [style=dot,  fill=red,  draw=red] (E1) at (2.9, 3.4) {};
		\node [style=none] (E2) at (3.4, 3.4) {\small{$(b, 1)$}};
		\node [style=none] (E) at (3.2, 3.2) {};
		\node [style=none] (E') at (3.2, 2.98) {};
		\node [style=none] (F) at (1.2, 1.55) {};
		\node [style=none] (F') at (1.2, 1.35) {};
		\node [style=dot,  fill=red,  draw=red] (F1) at (1, 1.1) {};
		\node [style=none] (F2) at (1.5, 1.1) {\small{$(c, 2)$}};
		\node [style=none] (H) at (1.4, 0.9) {};
		\node [style=none] (H') at (1.4, 0.7) {};
		\node [style=none] (G) at (3.2, 1.55) {};
		\node [style=none] (G') at (3.2, 1.35) {};
		\node [style=dot,  fill=red,  draw=red] (G1) at (2.7, 1.1) {};
		\node [style=none] (G2) at (3.2, 1.1) {\small{$(c, 2)$}};
		\node [style=none] (I) at (3.0, 0.9) {};
		\node [style=none] (I') at (3.0, 0.7) {};
		\node [style=none] (J) at (2.2,-1.22) {};
		\node [style=none] (J') at (2.2,-1.5) {};
		\node [style=dot,  fill=red,  draw=red] (J1) at (1.7, -1.7) {};
		\node [style=dot,  fill=red,  draw=red] (J2) at (1.92, -1.7) {};
		\node [style=none] (J3) at (3.0, -1.7) {\small{$((d,d), (4,4))$}};
	\end{pgfonlayer}
	\begin{pgfonlayer}{edgelayer}
		%
		\draw [style=simple]  (a) to (b);
		\draw [style=simple]  (a) to (c);
		\draw [style=simple]  (b) to (c);
		\draw [style=simple]  (d) to (e);
		\draw [style=simple]  (d) to (f);
		\draw [style=simple]  (e) to (f);
		\draw [style=simple]  (h) to (i);
		\draw [style=simple]  (h) to (j);
		\draw [style=simple]  (i) to (j);
		\draw [style=simple] (D) to (D');
		\draw [style=simple] (E) to (E');
		\draw [style=simple] (F) to (F');
		\draw [style=simple] (G) to (G');
		\draw [style=simple] (H) to (H');
		\draw [style=simple] (I) to (I');
		\draw [style=simple] (J') to (J);
		\draw [style=simple, draw=red] (tau1) to (Atau1);
		\draw [style=simple, draw=red] (tau1) to (Ctau1);
		\draw [style=simple, draw=red] (tau2) to (Btau2);
		\draw [style=simple, draw=red] (tau2) to (Ctau2);
		\draw [style=simple, draw=red] (tau4) to (C2tau4);
		\draw [style=simple, draw=red] (tau4) to (C1tau4);
		\draw [style=simple, draw=red] (tau4) to (D2tau4);
		\draw [style=simple, draw=red] (tau4) to (D1tau4);
	\end{pgfonlayer}
\end{tikzpicture}
}
\caption{Coordinate tasks to compose}
\vspace*{-4pt}
  \label{fig:exTaskOp}
     \end{subfigure}
  \caption{Primitive operations are composed for two UH60s ({\textcolor{red}{$\bullet$}}) to rendezvous at $c$ and maneuver together to $d$. Each primitive operation is indexed by a transition; types and space-time points must match to compose.}
  \label{fig:TaskingOp}
  \vspace*{-5pt}
\end{figure}

\subsection{Synthesis}
\label{subsec:netOPIntro}

The third example describes the automated design of mission task plans for SAR using network operads.  
The SAR tasking example illustrates the expressive power of applying existing operads and their potential to streamline and automate design synthesis. Fig.\ \ref{fig:PetriInputs} is analogous to  Fig.\ \ref{fig:sailnetop}, but whereas a sparse matrix specify an architecture problem, here a Petri net is used to model coordinated groups of agents. 

For the SAR tasking problem, much of the complexity results from agents' need to coordinate in space and time--e.g.\ when a helicopter is refueled in the air, as in $\tau_3$ of Fig.\ \ref{fig:PetriInputs}.  To facilitate coordination, the types of the network operad are systematically extended via a network model whose target categories add space and time dimensions; compare, e.g., \cite{NMPetri}.    
In this way, task plans are constrained at the level of syntax to enforce these key coordination constraints; see, e.g., Fig.\ \ref{fig:exTaskOp} where two UH60s at the same space-time point ({\textcolor{red}{$\bullet$}} $(c, 2)$) maneuver together to $d$.  We describe automated synthesis for this example in Sec.\ \ref{sec:auto}.

 \section{Cookbook modeling of domain problems}
 \label{sec:cookbook}
  In this section we describe some techniques for constructing operads and their algebras, using an example-driven, cookbook-style approach.  We emphasize recent developments for network operads and dive deeper into the SAR architecture problem.
\subsection{Network models}
\label{subsec:cooknetmod}
The theory of network models provides a general method to construct an operad $\OO$ by mixing combinatorial and compositional structures. Note that this lives one level of abstraction \emph{above} operads; we are interested in \emph{constructing} a language to model systems--e.g. for a specific domain.  This provides a powerful alternative to coming up with operads one-by-one.
A general construction allows the applied practitioner to cook-up a domain-specific syntax to compose systems by specifying some combinatorial ingredients.  

The first step is to specify what the networks to be composed by $\OO$ look like. Often this is some sort of graph, but what kind? Are nodes typed (e.g., colored)? Are edges symmetric or directed? Are loops or parallel edges allowed? What about $n$-way relationships for $n>2$ (hyper-edges)? 
We can mix, match and combine such combinatorial data to define different \emph{network models}, which specify the  system types and kinds of relationships between them relevant to some domain problem.  
The network model describes the operations 
we need to compose the networks specific to the problem at hand.

Three compositional structures which describe the algebra of operations. The \emph{disjoint} or \emph{parallel} structure combines two operations for networks with $m$ and $n$ nodes, respectively, into a single operation for networks with $m+n$ nodes. More restrictively, the \emph{overlay} or \emph{in series} structure superimposes two operations to design networks on $n$ nodes. The former structure combines separate operations to support modular development of designs; the latter supports an incremental design process, either on top of existing designs or from scratch.
The last ingredient permutes nodes in a network, which assures coherence between different ordering of the nodes. 
This last structure is often straightforward to specify.  If it is not, one should consider if symmetry is being respected in a natural way.  

We can distill the main idea behind overlay by asking, what happens when we add an edge to a network? It depends on the kind of network being  composed by $\OO$: 
\vspace*{-7pt}
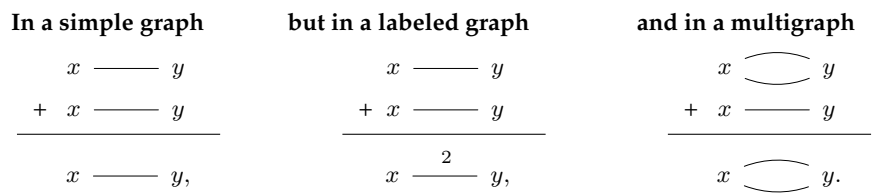
\begin{center}
\begin{tabular}{rrrrrr}
 \multicolumn{2}{r}{\textbf{In a simple graph}}& \multicolumn{2}{r}{\textbf{but in a labeled graph}}& 
	\multicolumn{2}{r}{\textbf{and in a multigraph}} 
	\\[1ex]
	\rule{0ex}{0ex}
	& $\xymatrix{x \ar@{-}[r] & y}\ $ 
	& \rule{8ex}{0ex}
	& $\xymatrix{x \ar@{-}[r] & y}\ \ $ 
	& \rule{8ex}{0ex}
	& $\xymatrix{x \ar@/^/@{-}[r] \ar@/_/@{-}[r] & y}\ $\\
	& + $\xymatrix{x \ar@{-}[r] & y}\ $ 
	&& +$\xymatrix{x \ar@{-}[r] & y}\ \ $ 
	&& + $\xymatrix{x \ar@{-}[r] & y}\ $\\ \cline{2-2}\cline{4-4}\cline{6-6}\\[-2.5ex]
	& $\xymatrix{x \ar@{-}[r] & y,}$ 
	&&$\xymatrix{x \ar@{-}[r]^{2} & y,}\ $ 
	& & $\xymatrix{x 
	\ar@/^/@{-}[r] 
	\ar@/_/@{-}[r]& y.}$
\end{tabular}
\vspace*{-5pt}
\end{center}
These differences are controlled by a \emph{monoid}\footnote{A set with a binary operation, usually written $\cdot$ unless the operation is commutative ($m+n=n+m$). A monoid is always associative, $\ell\cdot(m\cdot n)=(\ell\cdot m)\cdot n$ and has a unit $e$ satisfying $e\cdot m=m=m\cdot e$--e.g.\ multiplication of $n\times n$ matrices.}, which provides each $+$ shown.  Above, the monoids are bitwise OR, addition, and maximum, respectively.  As a further example, if edge addition is controlled by $\mathbb{Z}/2\mathbb{Z}$ then $+$ will have a toggling effect.

\begin{figure}[!ht]
 \centering 
\scalebox{0.9}{
\begin{tikzpicture}
	\begin{pgfonlayer}{nodelayer}
		\node [style=none] (LD) at (1.5, 2.75) {};
		\node [above] (LD') at (1.5, 3) {$2$};
		\node [above] (RE') at (2.8, 3) {$2$};
	    \node [style=dot] (L1) at (1.5, 2.6) {};
	    \node [style=dot] (L2) at (1.5, 2.25) {};
		\node [style=none] (La) at (1,2.8) {};
		\node [style=none] (Lb) at (2,2.8) {};
		\node [style=none] (Lc) at (1.5,1.9) {};
		\node [style=none] (Lc') at (1.5,1.95) {};
        \node [style=none] (RE) at (2.8, 2.75) {};
        \node [above] (RE') at (2.8, 3) {$2$};
	    \node [style=dot] (R1) at (2.8, 2.6) {};
	    \node [style=dot] (R2) at (2.8, 2.25) {};
	  	\node [style=none] (Ra) at (2.3,2.8) {};
		\node [style=none] (Rb) at (3.3,2.8) {};
		\node [style=none] (Rc) at (2.8,1.9) {};
		\node [style=none] (Rc') at (2.8,1.95) {};
		\node [style=dot] (1) at (1.95, 1) {};
		\node [style=dot] (2) at (2.45, 1) {};
		\node [style=dot] (3) at (1.95, 0.6) {};
		\node [style=dot] (4) at (2.45, 0.6) {};
		\node [style=none] (a) at (1,1.2) {};
		\node [style=none] (b) at (3.4,1.2) {};
		\node [style=none] (c) at (2.2,0) {};
		\node [style=none] (c') at (2.2,0.05) {};
		\node [style=none] (D) at (1.5, 1.15) {};
		\node [style=none] (E) at (2.8, 1.15) {};
		\node [style=none, outer sep =3pt] (D') at (1.5, 1.55) {$2$};
		\node [style=none, outer sep =3pt] (E') at (2.8, 1.55) {$2$};
		\node [below] (C) at (2.2, -0.2) {$4$};
		\node [style={font=\huge}] at (4.2, 1.8) {$\sqcup$};
		\node [style={font=\huge}] at (4.2, 1.2) {$\mapsto$};
	%
	    \node [style=none] (MD) at (5.5, 2.75) {};
		\node [above] (MD') at (5.5, 3) {$2$};
		\node [style=dot] (ML1) at (5.9, 2.6) {};
	    \node [style=dot] (ML2) at (5.9, 2.3) {};
	    \node [style=dot] (MR1) at (6.5, 2.6) {};
	    \node [style=dot] (MR2) at (6.5, 2.3) {};
	   \node [style=none] (ME) at (6.8, 2.75) {};
        \node [above] (ME') at (6.8, 3) {$2$};
	    \node [style=none] (UMa) at (5,2.8) {};
		\node [style=none] (UMb) at (7.4,2.8) {};
		\node [style=none] (UMc) at (6.2,1.9) {};
		\node [style=none] (UMc') at (6.2,1.95) {};
		\node [style=dot] (M1) at (5.95, 1) {};
		\node [style=dot] (M2) at (6.45, 1) {};
		\node [style=dot] (M3) at (5.95, 0.6) {};
		\node [style=dot] (M4) at (6.45, 0.6) {};
		\node [style=none] (Ma) at (5,1.2) {};
		\node [style=none] (Mb) at (7.4,1.2) {};
		\node [style=none] (Mc) at (6.2,0) {};
 		\node [style=none] (Mc') at (6.2,0.05) {};
 		\node [style=none] (MM) at (6.2, 1.15) {};
 		\node [style=none, outer sep =3pt] (MM') at (6.2, 1.55) {$4$};
 		\node [below] (MC) at (6.2, -0.2) {$4$};
		\node [style={font=\huge}] at (8.2, 1.8) {$+$};
		\node [style={font=\huge}] at (8.2, 1.2) {$\mapsto$};
	%
	    \node [style=none] (FD) at (9.5, 2.25) {};
		\node [above] (FD') at (9.5, 2.5) {$2$};
        \node [style=none] (FE) at (10.8, 2.25) {};
        \node [above] (FE') at (10.8, 2.5) {$2$};
		\node [style=dot] (F1) at (9.95, 1.9) {};
		\node [style=dot] (F2) at (10.45, 1.9) {};
		\node [style=dot] (F3) at (9.95, 1.4) {};
		\node [style=dot] (F4) at (10.45, 1.4) {};
		\node [style=none] (Fa) at (9,2.3) {};
		\node [style=none] (Fb) at (11.4,2.3) {};
		\node [style=none] (Fc) at (10.2,0.3) {};
 		\node [style=none] (Fc') at (10.2,0.35) {};
 		\node [below] (FC) at (10.2, 0) {$4$};
	\end{pgfonlayer}
	\begin{pgfonlayer}{edgelayer}
	    \draw [style=simple] (LD) to (LD');
	    \draw [style=simple] (L1) to (L2);
		\draw [style=simple] (La) to (Lb);
		\draw [style=simple] (La) to (Lc);
		\draw [style=simple] (Lb) to (Lc);
		\draw [style=simple] (Lc') to (D');
		\draw [style=simple] (Rc') to (E');
		\draw [style=simple] (RE) to (RE');
	    \draw [style=simple] (R1) to (R2);
		\draw [style=simple] (Ra) to (Rb);
		\draw [style=simple] (Ra) to (Rc);
		\draw [style=simple] (Rb) to (Rc);
		\draw [style=simple] (1) to (2);
		\draw [style=simple] (3) to (4);
		\draw [style=simple] (a) to (b);
		\draw [style=simple] (a) to (c);
		\draw [style=simple] (b) to (c);
		\draw [style=simple] (D) to (D');
		\draw [style=simple] (E) to (E');
		\draw [style=simple] (c') to (C);
	    \draw [style=simple] (MD) to (MD');
	    \draw [style=simple] (ME) to (ME');
	    \draw [style=simple] (ML1) to (ML2);
	    \draw [style=simple] (MR1) to (MR2);
	    \draw [style=simple] (UMa) to (UMb);
		\draw [style=simple] (UMa) to (UMc);
		\draw [style=simple] (UMb) to (UMc);
		\draw [style=simple] (UMc') to (MM');
		\draw [style=simple] (M1) to (M2);
		\draw [style=simple] (M3) to (M4);
	    \draw [style=simple] (Ma) to (Mb);
		\draw [style=simple] (Ma) to (Mc);
		\draw [style=simple] (Mb) to (Mc);
		\draw [style=simple] (MM) to (MM');
		\draw [style=simple] (Mc') to (MC);
		%
		\draw [style=simple] (FD) to (FD');
		\draw [style=simple] (FE) to (FE');
		\draw [style=simple] (F1) to (F2);
		\draw [style=simple] (F3) to (F4);
		\draw [style=simple] (F1) to (F3);
		\draw [style=simple] (F2) to (F4);
	    \draw [style=simple] (Fa) to (Fb);
		\draw [style=simple] (Fa) to (Fc);
		\draw [style=simple] (Fb) to (Fc);
		\draw [style=simple] (Fc') to (FC);
	\end{pgfonlayer}
\end{tikzpicture}
}
 \caption{Parallel ($\sqcup$) and in series ($+$) compositional structures define how to combine operations.}
\label{fig:comp}
\vspace*{-5pt}
 \end{figure}
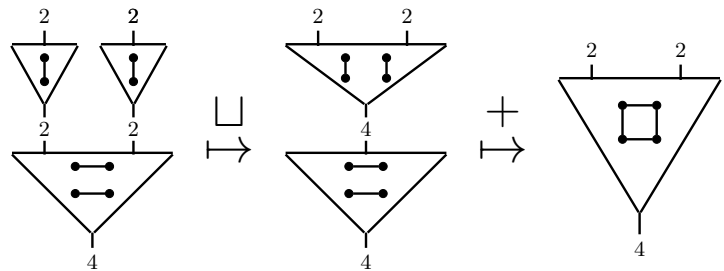

Consider simple graphs. Given a set of nodes $\n$, write $U_\n$ for the set of all undirected pairs $i \neq j$ (a.k.a.\ simple edges $i\mhyphen j$), so that $|U_\n|={\binom{|\n|}{2}}$. Then we can represent a simple graph over $\n$ as a $U_\n$-indexed vector of bits $\<b_{i\mhyphen j}\>$ describing which edges to `turn on' for a design. 
Each bit defines whether or not to add an $i\mhyphen j$ edge to the network and the overlay compositional structure is given by the monoid $\SG(\n):=\Bit^{U_\n}$, whose $+$ is bitwise OR for the product over simple edges--i.e.\ adding $i\mhyphen j$ then adding $i\mhyphen j$ is the same as adding $i\mhyphen j$ a single time. The disjoint structure
$\sqcup:\SG(\m)\times \SG(\n) \longrightarrow \SG(\mplusn)$ forms the disjoint sum of the graphs $g$ and $h$. Finally, permutations act by permuting the nodes of a simple graph. 
Together, these compositional structures define a network model  $\SG \maps \S \to \Mon$ which determines how operations are composed in the constructed network operad; see, Fig.\ \ref{fig:comp} or \cite[3.2, 7.2]{NetworkModels} for complete technical details.
 
This definition has an analogue for $\N$-weighted graphs, $\LG(\n):=(\N, + )^{U_\n}$, with overlay given by sum of edge weights and another for multi-graphs, $\MG(\n):=(\N, \max )^{U_\n}$, with overlay equivalent to union of multisets;
see  \cite[3.3, 3.4]{NetworkModels} for details.
More generally, we can label edges with the elements of \emph{any} monoid. Many of these examples are strange---binary addition makes edges cancel when they add---but their formal construction is straightforward; see \cite[Thm.\ 3.1]{NetworkModels}.

Equivalently, we can view the undirected edges in $U_\n$ as generators, subject to certain idempotence and commutativity relations: 
$
\SG(\n):= \langle e \in U_\n | e\cdot e = e, e\cdot e' = e' \cdot e \rangle.$
Here the idempotence relations come from $\Bit$ while the commutativity relations promote the single copies of $\Bit$ for each $i\mhyphen j$ to a well-defined network model.
Similar tricks work for lots of other network templates; we just change the set of generators to allow for new relationships.
For example, to allow self-loops, we add loop edge  generators $L_\n=\n+U_\n$ to express
relationships from a node $i$ to itself.
Likewise, network operads for directed graphs
can be constructed by using generators $D_\n=\n \times \n$, and one can also introduce higher-arity relationships. 

In all cases, the formal definition of a network model assures that all the combinatorial and compositional ingredients work well together;
one precise statement of ``working well together'' is given in \cite[2.3]{NetworkModels}.
Once a \emph{network template}---which expresses minimal data to declare the ingredients for a network model---is codified in a theorem 
as in \cite[3.1]{NetworkModels}, it can be reused in a wide variety of domains to set up the specifics of composition.

 \subsection{Cooking with operads}
 \label{subsec:cookop}
 The prototype for network operads is a simple network operad, which models only one kind of thing, such as aircraft.
 The types of a simple network operad are natural numbers, which serve to indicate how many aircraft are in a design. Operations of the simple network operad are simple graphs on some number of vertices.  
 For example, Fig.\ \ref{fig:comp} above shows a simple network operad to describe a design for point-to-point communication between aircraft.
 
Structural network operads extend this prototype in two directions: (1)
a greater diversity of things-to-be-modeled is supported by an expanded collection of types; and
    (2) more sorts of links or relationships between things are expressed via operations.
To illustrate the impact of network templates, 
suppose we are  
modeling heterogeneous system types with multiple kinds interactions.  For simplicity we consider simple interactions, which can be undirected or directed. 

A \emph{network template} need only declare the \emph{primitive} ways system types can interact to define a network model--e.g.\ a list of tuples $\textrm{(directed : carrying, } \HELO, \textrm{ } \CUTTER\textrm{)}$.   
This data is minimal in two ways: (1) \emph{any} framework must provide data to specify potentially valid interactions; and (2) this approach allows \emph{only} those interactions that make sense upon looking at the types of the systems involved.  Thus, interactions must be syntactically correct when constructing system designs.
 
  Presently, we will consider an example from the DARPA CASCADE program: the sailboat problem introduced in \ref{sec:examples}\ref{subsec:exSailboat}.
  This SAR application problem was inspired by the 1979 Fastnet Race and the 1998 Sydney to Hobart Yacht Race, in which severe weather conditions resulted in many damaged vessels distributed over a large area.
  Both events were tragic, with 19 and 6 deaths, respectively, and remain beyond the scale of current search and rescue planning.  
  Various larger assets---e.g.\ ships, airplanes, helicopters---could be based at ports and ferry smaller search and rescue units---e.g.\ small boats, quadcopters---to the search area.  
  
  Specifically, there were 8 atomic types to model:
  $
  P = \{ \PORT, \CUTTER, \BOAT, \FWS, \FWSAR, \HELO, \FWUAV, \QUAD  \}.
  $
  The primary relationship to specify a structural design is various assets carrying another types, so only one kind of interaction is needed: carrying.  
  This relationship is directed; e.g.,  a cutter ($\CUTTER$) can carry a helicopter ($\HELO$)  but not the other way around.
 Specifying allowed relationships amounts to specifying pairs of type $(p,p')\in  P \times P$ such that type $p'$ can carry type $p$;  see Fig.\ \ref{fig:sailnetop} for examples. 
 Fig.\ \ref{fig:sailnetop} data is extended to: (1) specify to that $\PORT$ can carry all types other than $\PORT$, $\FWUAV$ and $\QUAD$; (2) conform to an input file format to declare simple directed or undirected interactions, e.g, the  JSON format in Fig.\ \ref{fig:sailnetopFile}.
 
  \begin{figure}[hb]
     \centering
     \begin{subfigure}[b]{0.6\textwidth}
         \centering
\scalebox{0.8}{
\fbox{%
    \parbox{\textwidth}{ \small
        \texttt{\{`colors' : [`port', `cut', \ldots , `qd'], \newline
         \hspace*{2mm}`directed' : \{ \\
         \hspace*{6mm} `carrying': \{ \\
         \hspace*{6mm} \hspace*{6mm} `cut': [`port'], \\
         \hspace*{6mm} \hspace*{6mm} `boat': [`port', `cut'], \\
        \hspace*{6mm} \hspace*{6mm} \ldots, \\
        \hspace*{6mm} \hspace*{6mm} `qd': [`cut', \ldots , `helo'] \} \} \} }
}
}
}
  \caption{Network template data to specify the operad ${\OO}_{Sail}$}
  \label{tab:synport2}
  \vspace*{-4pt}
     \end{subfigure}
     \hfill
     \begin{subfigure}[b]{0.32\textwidth}
         \centering
         {
\begin{tikzpicture}
	\begin{pgfonlayer}{nodelayer}
		\node [style=dot, draw=ForestGreen, fill=ForestGreen] (1) at (2, 1) {}; 
		\node [style=dot, draw=red, fill=red] (2) at (2.4, 1) {}; 
		\node [style=dot, draw=blue, fill=blue] (3) at (2, 0.6) {}; 
		\node [style=dot, draw=blue, fill=blue] (4) at (2.4, 0.6) {};
		\node [style=none] (a) at (1.4,1.2) {};
		\node [style=none] (b) at (3.0,1.2) {};
		\node [style=none] (c) at (2.2,-.2) {};
		\node [style=dot, draw=ForestGreen, fill=ForestGreen] (D1) at (1.7, 1.6) {};
		\node [style=dot, draw=blue, fill=blue] (D2) at (1.9, 1.6) {};
		\node [style=none] (D) at (1.8, 1.15) {};
		\node [style=none] (D') at (1.8, 1.5) {};
		\node [style=dot, draw=red, fill=red] (E1) at (2.5, 1.6) {};
		\node [style=dot, draw=blue, fill=blue] (E2) at (2.7, 1.6) {};
		\node [style=none] (E) at (2.6, 1.15) {};
		\node [style=none] (E') at (2.6, 1.5) {};
		\node [style=none] (c') at (2.2,-.15) {};
		\node [style=none] (C) at (2.2, -0.5) {};
		\node [style=dot, draw=ForestGreen, fill=ForestGreen] (C1) at (2.1, -0.6) {};
		\node [style=dot, draw=red, fill=red] (C2) at (2.3, -0.6) {};
		\node [style=dot, draw=blue, fill=blue] (C3) at (2.1, -0.8) {};
		\node [style=dot, draw=blue, fill=blue] (C4) at (2.3, -0.8) {};
		
		\node [style=none] (Cut) at (3.92, 0.8) {{\textcolor{ForestGreen}{$\bullet$}} $\CUTTER$};
		\node [style=none] (Helo) at (4, 0.4) {{\textcolor{red}{$\bullet$}} $\HELO$};
		\node [style=none] (QD) at (3.845, -0.02) {{\textcolor{blue}{$\bullet$}} $\QUAD$};
	\end{pgfonlayer}
	\begin{pgfonlayer}{edgelayer}
		\draw [->, semithick] (3) to (1);
		\draw [->, semithick] (4) to (2);
		\draw [style=simple]  (a) to (b);
		\draw [style=simple]  (a) to (c);
		\draw [style=simple]  (b) to (c);
		\draw [style=simple] (D) to (D');
		\draw [style=simple] (E) to (E');
		\draw [style=simple] (c') to (C);
	\end{pgfonlayer}
\end{tikzpicture}
}
\caption{Example operation $\fop \in {\OO}_{Sail}$}
  \label{fig:exSailboat2}
  \vspace*{-4pt}
     \end{subfigure}
         \caption{After specifying ${\OO}_{Sail}$, $\fop$ places a  $\QUAD$ ({\textcolor{blue}{$\bullet$}}) on a $\CUTTER$ ({\textcolor{ForestGreen}{$\bullet$}}) and another $\QUAD$ ({\textcolor{blue}{$\bullet$}}) on a $\HELO$ ({\textcolor{red}{$\bullet$}}).}
\label{fig:sailnetopFile}
\end{figure}

If another type of system or kind of interaction is needed, then the file is appropriately extended. 
For example, we can include buoys by appending {\tt Buoy} to the array of {\tt colors} and augmenting the relationships in the {\tt carrying} node.  
Or, we can model the undirected (symmetric) relationship of communication by including an entry such as {\tt `undirected': \{`communication': \{`port`: [`cut', ...], ...\}\}}.
Moreover, modifications to network templates--such as ignoring $\textrm{(undirected : communication)}$ or combining $\QUAD$ and $\FWUAV$ into a single type--naturally induce mappings between the associated operads \cite[5.8]{NetworkModels}.     
 \subsection{Cooking with algebras}
 \label{subsec:cookalg}
 Because all designs are generated from primitive operations to add edges,  it is sufficient to define how primitive operations act in order to define an algebra. 
 For the sailboat problem, semantics are oriented to enable the delivery of a high capacity for search---known in the literature as search effort \cite[3.1]{Moving}---in a timely manner.  Given key parameters for each asset--e.g.\ speed, endurance, search efficiency across kinds of target and conditions, parent platform, initial locations--and descriptions of the search environment--e.g.\ expected search distribution, its approximate evolution over time--the expected number of surviving crew members found by the system can be estimated \cite[Ch.\ 3]{Moving}.

 Among these data, the parent platform and initial locations vary within a scenario and the rest describe the semantics of a given scenario.  In fact, we assume all platforms must trace their geographical location to
 one of a small number of base locations, so that the system responds from bases, but is organized to support rapid search. 
 Once bases are selected, the decision problem 
 is a choice of operation: what to bring (type of the composite system) and how to organize it (operation to carry atomic systems).  Data for the operational context specifies a particular algebra; see, e.g., Table \ref{tab:properties}.  
 Just as for the operad, this data is lightweight and configurable.  
 \begin{table}[t]
  \caption{Example properties captured in algebra for sailboat problem including time on station (ToS), speed for search (S) and max speed (R), and sweep widths measuring search efficiency for target types person in water (PIW), crew in raft (CIR) and demasted sailboats (DS) adrift. }
\begin{center}
 \begin{tabular}{ p{1.4cm} p{1.4cm} p{1.4cm} p{2cm} p{3.2cm}}
 \hline Type 
 & Cost (\$) 
 & ToS (hr)
 &
  \begin{tabular}{ p{0.7cm} p{0.7cm} }
                  \multicolumn{2}{c}{Speed (kn)}
                  \\
                  S & R\\
\end{tabular}
 &
   \begin{tabular}{ p{0.7cm} p{0.7cm} p{0.7cm} }
                  \multicolumn{3}{c}{Sweep Width (nmi)}
                  \\
                  PIW & CIR & DS \\ \end{tabular}
 \\
 \hline
 $\CUTTER$ & 200M & $\infty$ & \begin{tabular}{p{0.7cm} p{0.7cm}}
                  11  & 28\\
                 \end{tabular}  & 
                 \begin{tabular}{p{0.7cm} p{0.7cm} p{0.7cm}}
                  0.5  & 4.7 & 8.5\\
                 \end{tabular}  \\
$\BOAT$ & 500K & 6 & 
 \begin{tabular}{p{0.7cm} p{0.7cm}}
                  22  & 35\\
                 \end{tabular} & 
                 \begin{tabular}{p{0.7cm} p{0.7cm} p{0.7cm}}
                  0.4  & 4.2 & 7.5\\
                 \end{tabular}   \\
$\FWS$ & 60M & 9 &  \begin{tabular}{p{0.7cm} p{0.7cm}}
                  180 &  220\\
                 \end{tabular} & 
                 \begin{tabular}{p{0.7cm} p{0.7cm} p{0.7cm}}
                  0.1  & 2.2 & 7.6\\
                 \end{tabular}  \\
$\FWSAR$ & 72M & 10 & \begin{tabular}{p{0.7cm} p{0.7cm}}
                  180 &  235\\
                 \end{tabular} & 
                 \begin{tabular}{p{0.7cm} p{0.7cm} p{0.7cm}}
                  0.5  & 12.1 & 16.6\\
                 \end{tabular}  \\  
$\HELO$ & 9M & 4 & \begin{tabular}{p{0.7cm} p{0.7cm}}
                  90 &  180\\
                 \end{tabular} & 
                 \begin{tabular}{p{0.7cm} p{0.7cm} p{0.7cm}}
                  0.5  & 1.5 & 4.8\\
                 \end{tabular}   \\
$\FWUAV$ & 250K & 3 & \begin{tabular}{p{0.7cm} p{0.7cm}}
                  30 &  45\\
                 \end{tabular} & 
                 \begin{tabular}{p{0.7cm} p{0.7cm} p{0.7cm}}
                  0.5  & 1.8 & 4.5\\
                 \end{tabular}  \\
$\QUAD$ & 15K & 4 & \begin{tabular}{p{0.7cm} p{0.7cm}}
                  35 &  52\\
                 \end{tabular} & 
                 \begin{tabular}{p{0.7cm} p{0.7cm} p{0.7cm}}
                  0.5  & 1.5 & 4.8\\
                 \end{tabular}   \\
\hline
 \end{tabular}
\label{tab:properties}
 \end{center}
 \end{table}
 
{\bf Related cookbook approaches.}
 \label{subsec:cookRelated}
 Though we emphasized network operads,  the generators 
 approach is often studied and lends itself to encoding such combinatorially data with a ``template,'' in a cookbook fashion.  
 The generators approach to ``wiring'' has been developed into a theory of hypergraph categories \cite{FongThesis, Hyper}, which induce wiring diagram operads.  
 Explicit presentations for various wiring diagram operads are given in \cite{OpWireYau}. Augmenting monoidal categories with combinatorially specified data has also been investigated, e.g.\ in \cite{Bells}.
 
 \section{Functorial Systems Analysis}
 \label{sec:anal}
 
  In this section we demonstrate the use of functorial semantics in systems analysis. As in \ref{sec:operads}\ref{subsec:mapmaker}, a functor establishes a relationship between a syntactic or combinatorial model of a system (components, architecture) and some computational refinement of that description. This provides a means to consider a given system from different perspectives, and also to relate those viewpoints to one another. To drive the discussion, we will focus on the Length Scale Interferometer (LSI) and its wiring diagram model introduced in \ref{sec:examples}\ref{subsec:wireOPIntro}.
  
  \subsection{Wiring diagrams}
\label{subsec:wireOPExamples}
 
Operads can be applied to organize both qualitative and quantitative descriptions of hierarchical systems. Because operations can be built up iteratively from simpler ones to specify a complete design, different ways to build up a given design provide distinct avenues for analysis.

Figure \ref{fig:decomps} shows a wiring diagram representation of a precision measurement instrument called the Length Scale Interferometer (LSI) designed and operated by the US National Institute of Standards and Technology (NIST). Object types are system or component boundaries; Fig.\ \ref{fig:decomps} has: 6 components, the exterior, and 4 interior boundaries. Each boundary has an interface specifying its possible interactions, which are implicit in Fig.\ \ref{fig:decomps}, but define explicit types in the operad.

An operation in this context represents one step in a hierarchical decomposition, as in \ref{sec:operads}\ref{subsec:tree_view}. For example, the blue boxes in Fig.\ \ref{fig:decomps} represent a functional decomposition of the LSI into length-measurement and temperature-regulation subsystems: $\varphiOp\maps\LengthSys,\TempSys\to\LSIob$. These are coupled via (the index of refraction of) a $\laser$ interaction and linked to interactions at the system boundary. The operation $\varphiOp$ specifies the connections between blue and black boundaries.   

Composition in a wiring diagram operad is defined by nesting. For this functional decomposition, two further decompositions $\lambdaOp$ and $\tauOp$ describe the components and interactions within $\LengthSys$ and $\TempSys$, respectively. The wiring diagram in Fig.\ \ref{fig:decomps} is the composite $\varphiOp(\lambdaOp,\tauOp)$.

This approach cleanly handles multiple decompositions. Here the red boxes define a second, control-theoretic decomposition $\kappaOp:\Sensors,\Actuators\to \LSIob$. Unsurprisingly, the system is tightly coupled from this viewpoint, with heat flow to maintain the desired temperature, mechanical action to modify the path of the laser, and a feedback loop to maintain the position of the optical focus based on measured intensity. The fact that these two viewpoints specify the \emph{same} system design is expressed by the equation: $\varphiOp(\lambdaOp,\tauOp)=\kappaOp(\sigmaOp,\alphaOp)$; see \ref{sec:operads}\ref{subsec:equate_view} for related discussion.

\subsection{A probabilistic functor}
\label{subsec:prob}
Wiring diagrams can be applied to document, organize and validate a wide variety of system-specific analytic models. Each model is codified as an algebra, a functor from syntax to semantics (\ref{sec:operads}\ref{subsec:mapmaker}). For the example of this section,  all models have the same source (syntax), indicating that we are considering the same system, but the target semantics vary by application.
We have already seen some functorial models: the algebras in \ref{sec:cookbook}\ref{subsec:cookalg}. These can be interpreted as functors from the carrying operad $\OO_{Sail}$ to the operad of sets and functions $\Set$. Though $\Set$ is the ``default'' target for operad algebras, there are many alternative semantic contexts tailored to different types of analysis. Here we target an operad of probabilities $\Prob$, providing a simple model of nondeterministic component failure.

The data for the functor is shown in Table \ref{tab:fail}. Model data is indexed by operations\footnote{Types and operations, more generally, but the types carry no data in this simple example.} in the domain, an operad $\WW$ extracted from the wiring diagram in Fig.~\ref{fig:decomps}. The functor assigns each operation to a probability distribution that specifies the chance of a failure in each subsystem, assuming some error within the super-system.
For example, the length measurement and temperature regulation subsystems are responsible for 40\% and 60\% of errors in the LSI, respectively. This defines a Bernoulli distribution $P_\varphiOp$. Similarly, the decomposition $\tauOp$ of the temperature system defines a categorical distribution with 3 outcomes: $\Box$, $\Bath$ and $\Lab$. 

Relative probabilities compose by multiplication. This allows us to compute more complex distributions for nested diagrams. For the operation shown in Fig.~\ref{fig:decomps}, this indicates that the bath leads to nearly half of all errors ($60\%\times 80\%=48\%$) in the system.

Operad equations must be preserved in the semantics. Since $\varphiOp(\lambdaOp,\tauOp)=\kappaOp(\sigmaOp,\alphaOp)$, failure probabilities of source components don't depend on whether we think of them in terms of functionality or control. For the bath,
this relative failure probability is
\[
	\overbrace{60\%}^{P_\varphiOp}\times\overbrace{80\%}^{P_\tauOp}=48\%=\overbrace{72\%}^{P_\kappaOp}\times \overbrace{66.7\%}^{P_\alphaOp},
\]
and five analogous equations hold for the other source components.

Functorial semantics separates concerns: different operad algebras answer different questions. Here we considered \emph{if} a component will fail.  The LSI example is developed further in \cite[
 4]{LSIPaper} by a second algebra describing \emph{how} a component might fail, with Boolean causal models to propagate failures. The two perspectives are complementary, and loc.\ cit.\ explores integrating them with algebra homomorphisms (\ref{sec:operads}\ref{subsec:mapmaker}).
\begin{table}[h!]
	\caption{Failure probabilities form an operad algebra for LSI component failure.}
	\centering
	\begin{tabular}{|c|ccc||c|ccc|}
		\hline
		\multirow{2}{*}{$P_\varphiOp$} & $ls$&$\mapsto$& 40\%
		& \multirow{2}{*}{$P_\kappaOp$} & $sn$ & $\mapsto$ & 28\%\\
		& $ts$&$\mapsto$& 60\%
		&& $ac$ & $\mapsto$ & 72\%
		\\\hline
		\multirow{3}{*}{$P_\lambdaOp$} & $in$ & $\mapsto$ & 10\%
		& \multirow{4}{*}{$P_\sigmaOp$} & $lb$ & $\mapsto$ & 21.4\%\\
		& $\opOp$ & $\mapsto$ & 30\%
		& & $bt$ & $\mapsto$ & 21.4\%\\
		& $ch$ & $\mapsto$ & 60\%
		& & $\opOp$ & $\mapsto$ & 42.9\%\\\cline{1-4}
		\multirow{3}{*}{$P_\tauOp$} & $ba$ & $\mapsto$ & 80\%
		& & $in$ & $\mapsto$ & 14.3\%\\\cline{5-8}
		& $bx$ & $\mapsto$ & 10\%
		& \multirow{2}{*}{$P_\alphaOp$} & $ch$ & $\mapsto$ & 33.3\%\\
		& $lb$ & $\mapsto$ & 10\%
		& & $ba$ & $\mapsto$ & 66.7\%\\\cline{1-8}
	\end{tabular}
	\label{tab:fail}
\end{table}

\subsection{Interacting semantics}
\label{subsec:LSI}

Its toy-example simplicity aside, the formulation of a failure model $\WW\to\Prob$, as in Table~\ref{tab:fail} is limited in at least two respects. First, it tells us \emph{which} components fail, but not \emph{how} or \emph{why}. Second, the model is static, but system diagnosis is nearly always a dynamic process. We give a high-level sketch of an extended analysis to illustrate the integration of overlapping functorial models. 

The first step is to characterize some additional information about the types in $\WW$ (i.e., system boundaries). We start with the dual notions of \emph{requirements} and \emph{failure modes}. For example, in the temperature regulation subsystem of the LSI we have
\[\begin{array}{ccc}
	T_{\laser} \leq 20.02\degC &\leftrightarrow& T_\laser\textsf{ too high} \\[.5ex]
	19.98\degC \leq T_\laser &\leftrightarrow& T_\laser\textsf{ too low} \\
	\vdots && \vdots
\end{array}\]

Requirements at different levels of decomposition are linked by traceability relations. These subsystem requirements trace up to the measurement uncertainty for the LSI as a whole. Dually, an out-of-band temperature at the subsystem level can be traced back to a bad measurement in the $\Box$ enclosure, a short in the $\Bath$ heater or fluctuations in the $\Lab$ environment.

Traceability is compositional: requirements decompose and failures bubble up. This defines an  operad algebra\footnote{Many operads are defined from ordinary categories using a symmetric monoidal products \cite[7]{Hermida}. If a category carries more than one product, we use a superscript to indicate which is in use. The the disjoint union (+) corresponds to the disjunctive composition ``a failure in one component \emph{or} another; soon we will use the Cartesian product $\times$ to consider the  conjunctive relationship between ``the state of one component \emph{and} the other''.} $\Req:\WW\to\Rel^+$. Functoriality expresses the composition of traceability requirement across levels. See \cite[5]{LSIPaper} discussion of how to link these relations with Table~\ref{tab:fail} data.

For dynamics, we need \emph{state}. We start with a state space for each interaction among components. For example, consider the $\laser$ interaction coupling $\Chassis$, $\Interfer$, and $\Box$.  The most relevant features of the laser are its vacuum wavelength $\lambda_0$ and the ambient temperature, pressure and humidity (needed to correct for refraction). This corresponds to a four-dimensional state-space (or a subset thereof)
\[
\State(\laser) \cong \overbrace{[-273.15,\infty)]}^{T_\laser}\times \overbrace{[0,\infty)]}^{P_\laser}\times \overbrace{[0,1]}^{RH_\laser}\times \overbrace{[0,\infty)}^{\lambda_0} \subseteq \RR^4.
\]
A larger product defines an \emph{external state space} at each system boundary
\[\begin{array}{rl}
\State(\TempSys)=&\State(\laser)\times\State(\temp)^2\times\State(\setPt)\times\State(\water)\\[1ex]
\State(\Box)=&\State(\laser)\times \State({\temp})\times\State(\heat)^2\\
\vdots\\
\end{array}\]

Similarly, we can define an \emph{internal state space} for each operation by taking the product over all the interactions that appear in that diagram. We can decompose the internal state space in terms of either the system boundary or the components\footnote{Coupled variables are formalized through a partial product called the pullback, a common generalization of the Cartesian product, subset intersection and inverse image constructions.}:
\[\begin{array}{rcl}
\State(\varphiOp)&\cong & \State(\LSIob)\times\overbrace{\State(\laser)}^{\mathclap{\textrm{hidden variable}}}\\[1ex]
&\cong & \State(\LengthSys)
\times \mathclap{\underbrace{\underset{\State(\laser)}{}}_{\textrm{coupled variable}}}
\State(\TempSys)
\end{array}\]
The projections from these (partial) products form a relation, and these compose to define a functor $\WW\to\Rel^{\times}$:
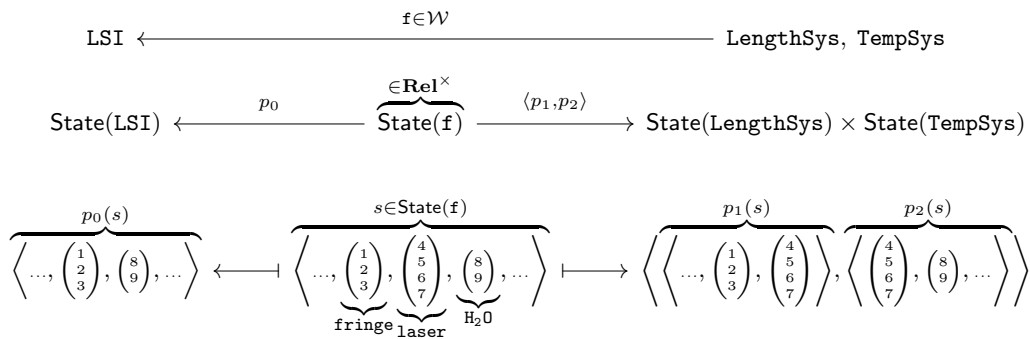
\[\xymatrix@R=1.5ex{
\LSIob && 
\LengthSys,\ \TempSys 
\ar[ll]_-{\varphiOp\in\WW} \\
\State(\LSIob) & {\overbrace{\State(\varphiOp)}^{\in\Rel^\times}} \ar[l]_-{p_0} \ar[r]^-{\<p_1,p_2\>} &
\State(\LengthSys)\times \State(\TempSys) \\
{\overbrace{\left\<\tdots,\cvect{1\\2\\3},\cvect{8\\9},\tdots\right\>}^{p_0(s)}}
&
{\overbrace{\left\< \tdots, \underbrace{\cvect{1\\2\\3}}_{\mathclap{\fringe}},
\underbrace{\cvect{4\\5\\6\\7}}_{\mathclap{\laser}},
\underbrace{\cvect{8\\9}}_{\mathclap{\water}},
\tdots\right\>}^{s\in\State(\varphiOp)}} \ar@{|->}[r] \ar@{|->}[l] & 
\left\<
{\overbrace{\left\<\tdots, \cvect{1\\2\\3}, \cvect{4\\5\\6\\7} \right\>}^{p_1(s)}},
{\overbrace{\left\< \cvect{4\\5\\6\\7}, \cvect{8\\9}, \tdots\right\>}^{p_2(s)}}
\right\> 
}\]

Each requirement $R\in \Req(\X)$ defines a subset $|R|\subseteq \State(\X)$, and a state is \emph{valid} if it satisfies all the requirements: $\Val(\X)=\bigcap_R |R|$. Using pullbacks (inverse image) we can translate validity to internal state spaces in two different ways. External validity (left square) checks that a system satisfies its contracts; joint validity (right square) couples component requirements to define the allowed joint states.
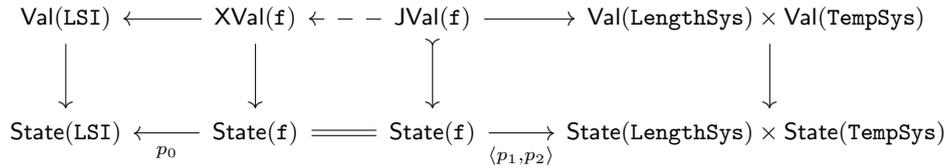
\[\vcenter{\xymatrix{
	\Val(\LSIob) \ar@{}[rrd]\ar[d] & \XVal(\varphiOp) \ar[l] \ar[d] & \JVal(\varphiOp) \ar@{-->}[l] \ar@{>->}[d] \ar[r] & \Val(\LengthSys)\times\Val(\TempSys) \ar@<1.2ex>[d] \\
	\State(\LSIob) & \State(\varphiOp) \ar[l]^-{p_0} \ar@{=}[r] & \State(\varphiOp) \ar[r]_-{\<p_1,p_2\>} & \State(\LengthSys)\times\State(\TempSys) \\
}}\]

A requirement model is \emph{sound} if joint validity entails external validity, corresponding to the dashed arrow above. With some work, one can show that these diagrams form the operations in an operad of entailments $\Ent$; see \cite[6]{CompAI} for a similar construction. The intuition is quite clear:
\[\begin{array}{crcl}
& \textrm{component reqs.} & \Rightarrow & \textrm{subsystem reqs.} \\
+ & \textrm{subsystem reqs.} & \Rightarrow & \textrm{system reqs.}\\\hline
& \textrm{component reqs.} & \Rightarrow & \textrm{system reqs.}\\
\end{array}\]

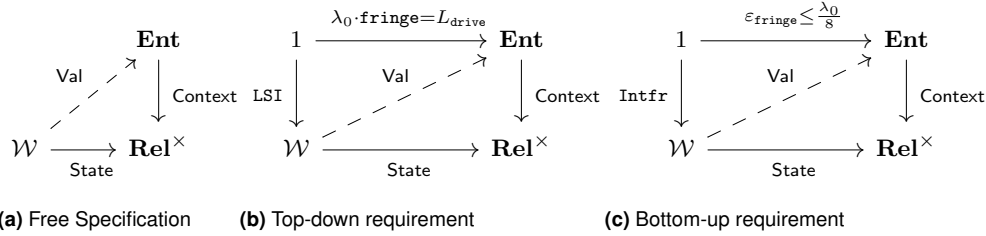
\begin{figure}
\centering
\begin{subfigure}[b]{0.23\textwidth}
\centering
$\xymatrix{
& \Ent \ar[d]^{\Context}\\
\WW \ar[r]_{\State} \ar@{-->}[ur]^{\Val} & \Rel^\times\\
}$
\caption{Free Specification}
\label{fig:liftsFree}
\vspace*{-10pt}
\end{subfigure}
\begin{subfigure}[b]{0.35\textwidth}
\centering
$\xymatrix{
1 \ar[d]_{\LSIob} \ar[rr]^{\lambda_0\cdot\fringe=L_\drive} && \Ent \ar[d]^{\Context}\\
\WW \ar[rr]_{\State} \ar@{-->}[urr]^{\Val} && \Rel^\times\\
}$
\caption{Top-down requirement}
\label{fig:liftsTop}
\vspace*{-10pt}
\end{subfigure}
\begin{subfigure}[b]{0.35\textwidth}
\centering
$\xymatrix{
1 \ar[d]_{\Interfer} \ar[rr]^{\varepsilon_{\fringe}\leq \frac{\lambda_0}{8}} && \Ent \ar[d]^{\Context}\\
\WW \ar[rr]_{\State} \ar@{-->}[urr]^{\Val} && \Rel^\times\\
}$
\caption{Bottom-up requirement}
\label{fig:liftsBot}
\vspace*{-10pt}
\end{subfigure}
\caption{Requirement specification expressed as lifting problems.}
\label{fig:lifts}
\end{figure}

There is a functor $\Context:\Ent\to\Rel^\times$, which extracts the relation across the bottom row of each entailment. Noting that the $\State$ relations occur in the validity entailment, we can reformulate requirement specification as a \emph{lifting problem} (Fig.~\ref{fig:liftsFree}): given functors $\State$ and $\Context$, find a factorization $\Val$ making the triangle commute. The second and third diagrams (Fig.~\ref{fig:liftsTop}--\ref{fig:liftsBot}) show how to extend the lifting problem with prior knowledge, in this case a top-level requirement and a known (e.g., off the shelf) component capability.

Finally we are ready to admit dynamics, but it turns out that we have already done most of the work. All that is needed is to modify the spaces attached to our interactions. In particular, we can distinguish between static and dynamic state variables; for the $\laser$,  $T$, $P$ and $RH$ are dynamic while $\lambda_0$ is static. Now we replace the static values $T,P,RH\in\RR$ by functions $T(t),P(t),RH(t)\in\RR^T$, thought of as \emph{trajectories} through the state space over a timeline $t\in\tau$. For example, we have 
\[
\Traj(\laser)\subseteq \overbrace{(\RR^\tau)^3}^{T,P,RH}\times\overbrace{\RR}^{\lambda_0}.
\]

From this, we construct $\Traj: \WW\to\Rel^\times$ using exactly the same recipe as above. Trajectories and states are related by a pair of algebra homomorphisms $\inst$ and $\const$. The first picks out a instantaneous state for each point in time, while the second identifies constant functions, which describe fixed-points of the dynamics:
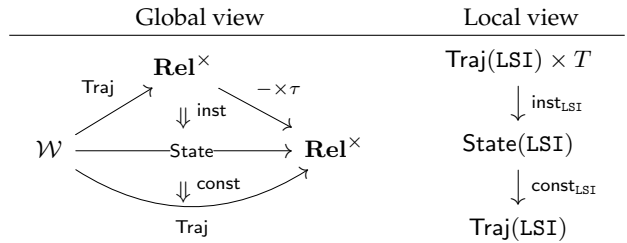
\[\begin{array}{ccc}
\textrm{Global view} && \textrm{Local view}\\\hline
\vcenter{\xymatrix@R=3ex{
& \Rel^\times \ar[rd]^{-\times \tau} \ar@{}[d]|(.6){\displaystyle\Downarrow}^{\inst}|(1.5){\displaystyle\Downarrow}^(1.4){\const}&\\
\WW \ar[ur]^{\Traj} \ar[rr]|{\State} \ar@/_5ex/[rr]_{\Traj}&& \Rel^\times
}} &&
\vcenter{\xymatrix@R=3ex{
\Traj(\LSIob)\times T \ar[d]^{\inst_{\LSIob}} \\
\State(\LSIob) \ar[d]^{\const_{\LSIob}}\\
\Traj(\LSIob)
}}
\end{array}
\]

The problem is that the state space explodes; function spaces are very large. Nonetheless, all of the system integration logic is identical, and using the entailment operad $\Ent$, we can build in additional restrictions to limit the search space. In particular, we can restrict attention to the subset of functions that satisfies a particular differential equation or state-transition relationship. This drastically limits the set of valid trajectories, though the resulting set may be difficult to characterize and the methods for exploring it will vary by context.

{\bf Related analytic applications.}
 \label{subsec:analyis}
Wiring diagrams have an established applied literature 
for system design problems, see, e.g., \cite{CyberWire, Seven, OpWire, CatSci,  SpivakTan, OpenDyn}.    
More broadly, the analytic strength of category theory to express compositionality and functional semantics 
is explored in numerous recent applied works--e.g.\ engineering diagrams \cite{ PaLin, Props,  OpenPetriNets, OpenCM, CoyaThesis, Seven, DigitalCircuits}, Markov processes \cite{CompMark, BioMark}, database integration \cite{BSW, Seven, CompPow, CatSci, SpivakKent, SpivakWis, MultiManu},  behavioral logic \cite{Seven, ToposSem, TempType, SheafEvent}, natural language processing \cite{FoundNLP, SentNPL, QuantNL}, machine learning \cite{Len, BackProp}, cybersecurity \cite{YonedaHack, SemanticsCyber, CyberWire}, quantum computation \cite{OptQuant, ReduceQuant, Quantomatic} and open games \cite{GameGraph, GameMixed, CompGame}.
 
\section{Automated synthesis with network operads}
\label{sec:auto}

An operad acting on an algebra provides a starting point to automatically generate and evaluate candidate designs.  Formally correct designs (operations in some operad) combine basic systems (elements of some algebra of that operad) into a composite system.

\subsection{Sailboat example}
\label{subsec:autoStruct}

Consider the sailboat problem introduced in  \ref{sec:examples}\ref{subsec:exSailboat} and
revisited in \ref{sec:cookbook}\ref{subsec:cookop}--\ref{subsec:cookalg}.
   Network operads describe assets and ports carrying each other while algebra-based semantics guided the search for effective designs by capturing the impact of available search effort. 
   To apply this model to automate design synthesis,
   algorithms explored designs within budget constraints based on costs in Table \ref{tab:properties}.
   Exploration iteratively composed up to budget constraints and operational limits on carrying\footnote{Though not used for this application, it turns of that degree limits--e.g.\ how many quadcopters a helicopter can carry--can be encoded directly into operad operations; the relevant mathematics was worked out in \cite{Moeller}.}.
   With these analytic models, greater sophistication was not needed; other combinatorial search algorithms--e.g.\ simulated annealing--are readily applied to large search spaces.
The most effective designs could ferry a large number of low cost search and rescue units--e.g.\ quadcopters ($\QUAD$)--quickly to the scene--e.g.\ via helicopters ($\HELO$).

\subsection{Tasking example}
\label{subsec:autoTask}
Surprisingly, network operads---originally developed to design systems---can also be applied to ``task" them: in other words, declare their behavior.  An elegant example of this approach is given in \cite{NMPetri} where ``catalyst" agents enable behavioral options for a system. 

{\bf The SAR tasking problem.}  
  The sailboat problem is limited by search: once sailboat crew members are found, their recovery is relatively straightforward. 
In hostile environments, recovery of isolated personnel (IPs) can become very complex. 
 The challenge is balancing the time criticality of recovery with the risk to the rescuers by judiciously orchestrating recovery teams\footnote{The recovery of downed airman Gene Hambleton, call sign Bat 21 Bravo, during the Vietnam War is a historical example of ill-fated SAR risk management. Hambleton's eventual  recovery cost 5 additional aircraft being shot down and 11 deaths; for comparison, a total 71 rescuers and 45 aircraft were lost to save 3,883 lives during Vietnam War SAR \cite{Bat21}. }.
 Consider the potential challenges of a large scale earthquake during severe drought conditions which precipitates multiple wildfires over a large area.
 The 2020 Creek Fire near Fresno, CA required multiple mass rescue operations (MROs) to rescue over 100 people in each case by pulling in National Guard, Navy and Marine assets to serve as search and rescue units (SRUs) \cite{FBsept8, FBsept6}.  
  Though MRO scenarios are actively considered by U.S. SAR organizations,
  the additional challenge of concurrent MROs distributed over a large area is not typically studied.  
  
  In this SAR tasking example, multiple, geographically distributed  IP groups compete for limited SRUs. 
  The potential of coordinating multiple agent types---e.g., fire fighting airplanes together with helicopters---to jointly overcome environment risks is considered as well as aerial refueling options for SRUs to extend their range.
  Depending on available assets, recovery demands and risks, a mission plan may need to work around some key agent types--e.g.\ refueling assets--and maximize the impact of others--e.g.\ moving protective assets between recovery teams.
  
  Under CASCADE, tasking operations were built up from primitive tasks that coordinate multiple agent types to form a composite task plan.  Novel concepts to coordinate teams of SRUs are readily modeled with full representation of the diversity of potential mission  plan solutions.
   
{\bf Network models for tasking.}  A network model for tasking defines atomic agent types $C$ and possible task plans for each 
list of agent types.  Whereas a network model to design structure $\mathsf{\Gamma} \maps \S(C) \to \Mon$  has values that are possible graphical designs, a network model to task behavior  $\mathsf{\Lambda} \maps \S(C) \to \Cat$ has values that are categories whose morphisms index possible task plans for the assembled types; compare, e.g., \cite[Thm.\ 9]{NMPetri}. Each morphism declares a sequence of tasks for each agent--many of which will be coordinated with other agents.

If the system is comprised of only a single UH-60 helicopter, its possible tasks are captured in $\mathsf{\Lambda}(\HH)$.  In this application, these tasks are paths in a graph describing `safe maneuvers.'
For unsafe maneuvers, UH-60s travel in pairs--or perhaps with escorts such as a HC-130 or CH-47 equipped with a Modular Airborne Fire Fighting System (MAFFS). Anything one UH-60 can do, so can two, but not vice versa. Thus there is a proper inclusion $
\mathsf{\Lambda}(\HH) \times \mathsf{\Lambda}(\HH) \subsetneq \mathsf{\Lambda}(\HH \otimes \HH)$.
\noindent  Similarly, $
\mathsf{\Lambda}(\HH) \times \mathsf{\Lambda}(\HC) \subsetneq \mathsf{\Lambda}(\HH \otimes \HC)$
\noindent since once both a UH-60 and HC-130 are present, a joint behavior of midair refueling of the UH-60 by the HC-130 becomes possible.  Formally, these inclusions are lax structure maps--e.g.\ $
\Phi_{(\HH, \HH)} \maps \mathsf{\Lambda}(\HH) \times \mathsf{\Lambda}(\HH) \to \mathsf{\Lambda}(\HH \otimes \HH)$,
specifies: given tasks for a single UH-60 (left coordinate) and tasks for another UH-60 (right coordinate), define the corresponding joint tasking of the pair. Here the joint tasking is: each UH-60 operates independently within the safe graph.  On the other hand, tasks in $\mathsf{\Lambda}(\HH \otimes \HH)$ to maneuver in unsafe regions can not be constructed from independent taskings of each UH-60.  Such tasks must be set for some pair or other allowed team--e.g.\ a CH-47 teamed with an UH-60. 

\begin{figure}[b]
\begin{subfigure}[b]{\textwidth}
 \centering
 \scalebox{0.95}{
\begin{tikzpicture}
	\begin{pgfonlayer}{nodelayer}
		\node [style=species] (A) at (-5, 0.8) {$\;\;a\;\;$};
		\node [style=species] (B) at (-5, -0.7) {$\;\;b\;\;$};
		\node [style=species] (C) at (-1, 0) {$\;\;c\;\;$};
		\node [style=species] (D) at (3, 0) {$\;\;d\;\;$};
		\node [style=transition] (tau1) at (-3, 0.6) {$\;\phantom{\Big{|}}\tau_1\;$};
		\node [style=transition] (tau2) at (-3, -0.6) {$\;\phantom{\Big{|}}\tau_2\;$};
        \node [style=transition] (tau3) at (-1, 1.8) {$\;\phantom{\Big{|}}\tau_3\;$};
        \node [style=transition] (tau4) at (1, 0) {$\;\phantom{\Big{|}}\tau_4\;$};
		\node [style=none] (HH) at (1.715, 1.8) {\small {\textcolor{red}{$\bullet$}} $\HH$};
		\node [style=none] (HC) at (1.8, 1.4) {\small {\textcolor{Orange}{$\bullet$}} $\HC$};
	\end{pgfonlayer}
	\begin{pgfonlayer}{edgelayer}
		\draw [style=inarrow, draw=red] (A) to [bend left=10] node[midway, above] {\tiny$\HH$} (tau1) ;
 		\draw [style=inarrow, draw=red] (tau1) to [bend left=10]  node[midway, above] {\tiny$\HH$} 
 		(C);
	    \draw [style=inarrow, draw=red] (B) to [bend right=10]  node[midway, below]{\tiny$\HH$  }  (tau2);
        \draw [style=inarrow, draw=red] (tau2) to [bend right=10]  node[midway, below]{\tiny$\HH$} (C); 
	    \draw [style=inarrow, draw=red] (C) to [bend right=20]  node[pos=0.7,right
	    ]{\tiny$\HH +\HC$} (tau3);
	    \draw [style=inarrow, draw=Orange] (C) to [bend right=10] (tau3);
        \draw [style=inarrow, draw=red] (tau3) to [bend right=20]  node[pos=0.3,left
	    ]{\tiny$\HH +\HC$} (C);
	    \draw [style=inarrow, draw=Orange] (tau3) to [bend right=10] (C);
	    \draw [style=inarrow, draw=red] (C)  to [bend left=5] node[midway, above]{\tiny$2\HH$} (tau4);
	    \draw [style=inarrow, draw=red] (C) to [bend right=5]  (tau4);
        \draw [style=inarrow, draw=red] (tau4) to [bend left=5] node[midway, above]{\tiny$2\HH$} (D);
        \draw [style=inarrow, draw=red] (tau4) to [bend right=5]  (D);
	\end{pgfonlayer}
\end{tikzpicture}
}
\vspace*{-4pt}
\caption{Four primitive tasks specified in a Petri net; arcs indicate types involved in each task.}
\vspace*{-10pt}
\label{fig:PTgen}
\end{subfigure}
\begin{subfigure}[b]{\textwidth}
         \centering
         {
\begin{tikzcd}[ampersand replacement=\&]
\scalebox{0.32}{
\begin{tikzpicture}
	\begin{pgfonlayer}{nodelayer}
		\node [style=species] (A) at (-5, 0.8) {$\;\;\;\;\;$};
		\node [style=species] (B) at (-5, -0.7) {$\;\;\;\;\;$};
		\node [style=species] (C) at (-1, 0) {$\;\;\;\;\;$};
		\node [style=species] (D) at (3, 0) {$\;\;\;\;\;$};
		\node [style=transition] (tau1) at (-3, 0.6) {$\;\;\phantom{\Big{|}}\;\;\;$};
		\node [style=transition] (tau2) at (-3, -0.6) {$\;\;\phantom{\Big{|}}\;\;\;$};
        \node [style=transition] (tau4) at (1, 0) {$\;\;\phantom{\Big{|}}\;\;\;$};
	\end{pgfonlayer}
	\begin{pgfonlayer}{edgelayer}
		\draw [style=inarrow, draw=red] (A) to [bend left=10] (tau1) ;
 		\draw [style=inarrow, draw=red] (tau1) to [bend left=10] (C);
	    \draw [style=inarrow, draw=red] (B) to [bend right=10]  (tau2);
        \draw [style=inarrow, draw=red] (tau2) to [bend right=10] (C); 
	    \draw [style=inarrow, draw=red] (C)  to [bend left=5]  (tau4);
	    \draw [style=inarrow, draw=red] (C) to [bend right=5]  (tau4);
        \draw [style=inarrow, draw=red] (tau4) to [bend left=5] (D);
        \draw [style=inarrow, draw=red] (tau4) to [bend right=5]  (D);
	\end{pgfonlayer}
\end{tikzpicture}
}
\&  \&
\arrow[d, Rightarrow, "\textrm{Network Model }"']
\arrow[rr]
\arrow[ll]
\scalebox{0.32}{
\begin{tikzpicture}
	\begin{pgfonlayer}{nodelayer}
		\node [style=species] (A) at (-5, 0.8) {$\;\;\;\;\;$};
		\node [style=species] (B) at (-5, -0.7) {$\;\;\;\;\;$};
		\node [style=species] (C) at (-1, 0) {$\;\;\;\;\;$};
		\node [style=species] (D) at (3, 0) {$\;\;\;\;\;$};
		\node [style=transition] (tau1) at (-3, 0.6) {$\;\;\phantom{\Big{|}}\;\;\;$};
		\node [style=transition] (tau2) at (-3, -0.6) {$\;\;\phantom{\Big{|}}\;\;\;$};
	\end{pgfonlayer}
	\begin{pgfonlayer}{edgelayer}
		\draw [style=inarrow, draw=red] (A) to [bend left=10] node[midway, above] {} (tau1) ;
 		\draw [style=inarrow, draw=red] (tau1) to [bend left=10]  node[midway, above] {} 
 		(C);
	    \draw [style=inarrow, draw=red] (B) to [bend right=10]   (tau2);
        \draw [style=inarrow, draw=red] (tau2) to [bend right=10]  (C); 
	\end{pgfonlayer}
\end{tikzpicture}
}
\&  \&
\scalebox{0.32}{
\begin{tikzpicture}
	\begin{pgfonlayer}{nodelayer}
		\node [style=species] (A) at (-5, 0.8) {$\;\;\;\;\;$};
		\node [style=species] (B) at (-5, -0.7) {$\;\;\;\;\;$};
		\node [style=species] (C) at (-1, 0) {$\;\;\;\;\;$};
		\node [style=species] (D) at (3, 0) {$\;\;\;\;\;$};
		\node [style=transition] (tau1) at (-3, 0.6) {$\;\;\phantom{\Big{|}}\;\;\;$};
		\node [style=transition] (tau2) at (-3, -0.6) {$\;\;\phantom{\Big{|}}\;\;\;$};
        \node [style=transition] (tau3) at (-1, 1.75) {$\;\;\phantom{\Big{|}}\;\;\;$};
	\end{pgfonlayer}
	\begin{pgfonlayer}{edgelayer}
		\draw [style=inarrow, draw=red] (A) to [bend left=10] (tau1) ;
 		\draw [style=inarrow, draw=red] (tau1) to [bend left=10]  
 		(C);
	    \draw [style=inarrow, draw=red] (B) to [bend right=10]   (tau2);
        \draw [style=inarrow, draw=red] (tau2) to [bend right=10] (C); 
	    \draw [style=inarrow, draw=red] (C) to [bend right=20]  (tau3);
	    \draw [style=inarrow, draw=Orange] (C) to [bend right=10] (tau3);
        \draw [style=inarrow, draw=red] (tau3) to [bend right=20] (C);
	    \draw [style=inarrow, draw=Orange] (tau3) to [bend right=10] (C);
	\end{pgfonlayer}
\end{tikzpicture}
}
 \\
 \mathsf{\Lambda}(\HH \otimes \HH)
 \&  \&
 \arrow[d, Rightarrow, "\textrm{Constraint Matrices  }"']
\arrow[rr]
\arrow[ll]
 \mathsf{\Lambda}(\HH)
 \&  \&
 \mathsf{\Lambda}(\HH \otimes \HC) \\
 {
 \begin{array}{c}
      M(\HH \otimes \HH) 
      \\
     \scalemath{0.5}
  {
\begin{bmatrix}
-1 &  0 &  1 &  0 &  0 &  0 &  0 &  0  \\
 0 & -1 &  1 &  0 &  0 &  0 &  0 &  0  \\
 0 &  0 &  0 &  0 & -1 &  0 &  1 &  0  \\
 0 &  0 &  0 &  0 &  0 & -1 &  1 &  0  \\
 0 &  0 & -1 &  1 &  0 &  0 & -1 &  1
\end{bmatrix}
}
\\
{}
\\
 M^s(\HH \otimes \HH) 
 \\
 \scalemath{0.5}
  {
\begin{bmatrix}
 1 &  0 &  0 &  0 &  0 &  0 &  0 &  0  \\
 0 &  1 &  0 &  0 &  0 &  0 &  0 &  0  \\
 0 &  0 &  0 &  0 &  1 &  0 &  0 &  0  \\
 0 &  0 &  0 &  0 &  0 &  1 &  0 &  0  \\
 0 &  0 &  1 &  0 &  0 &  0 &  1 &  0
\end{bmatrix}
 }
 \end{array}
 }
\&  \& 
 {
 \begin{array}{c}
 M(\HH) \\
     \scalemath{1.0}
  {
\begin{bmatrix}
-1 &  0 &  1 & 0  \\
 0 & -1 &  1 & 0
\end{bmatrix}
}
\\
{}
\\
M^s(\HH) 
\\
     \scalemath{1.0}
  {
\begin{bmatrix}
1 &  0 &  0 & 0  \\
 0 & 1 &  0 & 0
\end{bmatrix}
}
 \end{array}
 }
 \&  \& 
{
 \begin{array}{c}
  M(\HH \otimes \HC)  \\
     \scalemath{0.75}
  {
\begin{bmatrix}
-1 &  0 &  1 &  0 &  0 &  0 &  0 &  0  \\
 0 & -1 &  1 &  0 &  0 &  0 &  0 &  0  \\
 0 &  0 &  0 &  0 &  0 &  0 &  0 &  0  \\
\end{bmatrix}
}
\\
{}
\\
  M^s(\HH \otimes \HC)  \\
     \scalemath{0.75}
  {
\begin{bmatrix}
 1 &  0 &  0 &  0 &  0 &  0 &  0 &  0  \\
 0 &  1 &  0 &  0 &  0 &  0 &  0 &  0  \\
 0 &  0 &  1 &  0 &  0 &  0 &  1 &  0  \\
\end{bmatrix}
}
 \end{array}
  } 
\end{tikzcd}
}
\vspace*{-4pt}
\caption{More primitive tasks become possible as available agent types increase.  Type update matrices $M(-)$ and target to source constraint matrices $M^s(-)$ translate type changing and matching, resp.}
\label{fig:PTconstraint}
\end{subfigure}
\vspace*{-16pt}
\caption{Specified primitive tasks determine an
operad $\OO_{SAR}$ and 
a constraint program to explore operations.}
\label{fig:PTTask}
\end{figure}
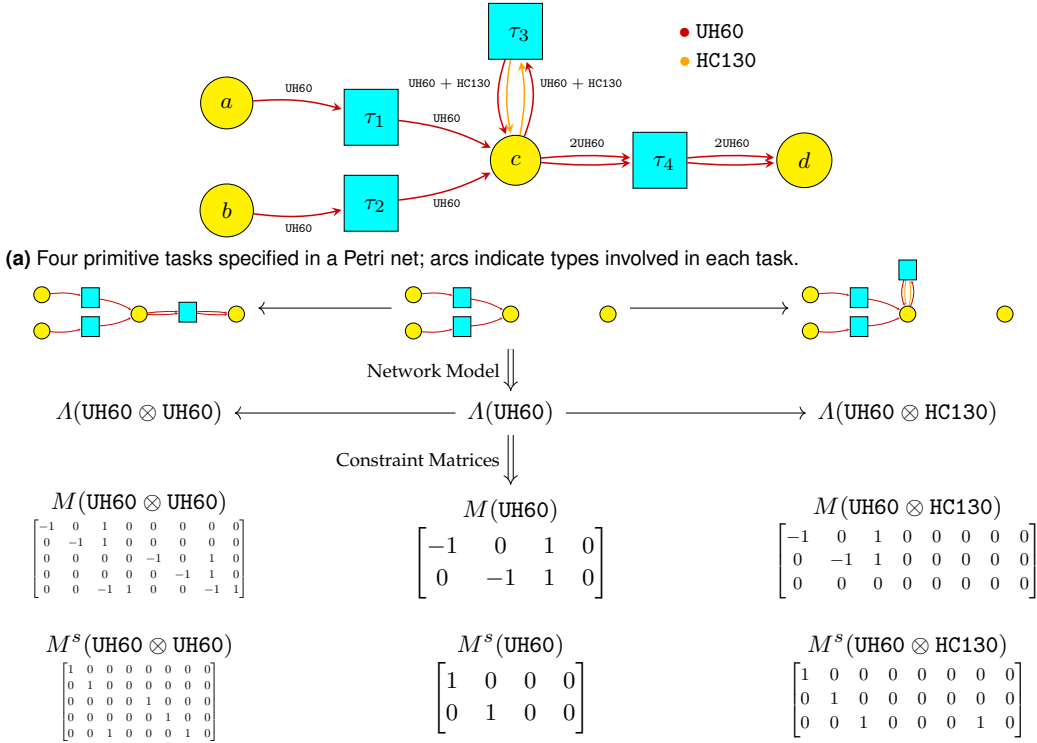

{\bf Applying the cookbook: operads.}
While the above discussion sketches how to specify a network model for tasking, which constructs a network operad \cite{NetworkModels}, these precise details \cite{ExpTasking} need not concern the applied practitioner\footnote{That is, a Petri net specifies the network model $\mathsf{\Lambda} \maps \S(C) \to \Cat$ to task behavior.  The construction of $\mathsf{\Lambda}$ \cite{ExpTasking} is similar to the construction described in \cite[Thm.\ 9]{NMPetri}, but adapted to colored Petri nets 
whose transitions preserve the number of tokens of each color; see, e.g., Fig.\ \ref{fig:PTgen}. Compared to \cite[Thm.\ 9]{NMPetri}, $C$ corresponds to token colors, rather than catalysts \cite[Def.\ 6]{NMPetri}, and species index discrete coordination locations.  Target categories encode allowed paths for each atomic agent type, (cont.)\\ $^9$ (cont.), e.g., for ~Fig.~\ref{fig:PTgen} $\mathsf{\Lambda}(\HH)$ is (freely) generated by objects $\{ a, b, c, d \}$ and morphisms $\tau_1\maps a \to c$ and  $\tau_2\maps a \to c$, whereas $\mathsf{\Lambda}(\HC)$ is just by generated $\{ a, b, c, d \}$ since no transition involves a single $\HC$.  By describing each target category as an appropriate subcategory of a product of path categories, the symmetric group action is given permuting coordinates, which allows the role of each atomic agent in a task to be specified.}. It is sufficient to provide a Petri net as template, from which a network operad is constructed.
Whereas a template to design structures defines the basic ways system types can interact, a template to task behavior defines the primitive tasks for agent types $C$, which are token colors in the Petri net.

No specification of `staying put' tasks are needed; these are implicit. All other primitive tasks are (sparsely) declared. For example, each edge of the `safe graph' for a solo UH-60 declares: (1) a single agent of type $\HH$ participates in this `traverse  edge' task; and
(2) participation is possible if a $\HH$ is available at the source of the edge. 
Likewise, each edge of the `unsafe graph' for pairs of UH-60s should declare similar information for pairs, but what about operations to refuel an UH-60 with a HC-130?
It turns out that transitions in a Petri net carry sufficient data \cite{NMPetri, ExpTasking} and have a successful history of specifying generators for a monoidal category \cite{OpenPetriNets, MM, CatNet}.
The Petri net Fig.\ \ref{fig:PTgen} shows examples where, for simplicity, tasks to traverse edges are only shown in the left to right direction.  
This sparse declaration is readily extended--e.g.\ to add recovery focused CH-47s, which tested their operational limits to rescue as many as 46 people during the 2020 Creek Fire--$C$ and the set of transitions are augmented to encode the new options for primitive tasks.

This specification of syntax is almost sufficient for the SAR tasking problem and would be for situations where only the sequence of tasks for each agent needs to be planned.  When tasking SAR agents, \emph{when} tasks are performed is semantically important because where and how long air-based agents `stay put' impacts success: (1) fuel burned varies dramatically for ground vs.\ air locations; (2) risk incurred varies dramatically for safe vs.\ unsafe locations. 
For comparison, in a ground-based domain without environmental costs, these considerations might be approximately invariant relative to the time tasks occur, and therefore, can be omitted from tasking syntax.   

Timing information creates little added burden for building a template--transitions declaring primitive tasks need only be given durations derivable from scenario data--and it is technically straightforward to add a time dimension to the network model.

{\bf Constraints from syntax.}
A direct translation of primitive tasks to decision variables for a constraint program is possible.  For syntax, the idea is very simple: enforce type matching constraints on composing operad morphisms. Here we will briefly indicate the original 
mixed integer linear program developed for SAR tasking; later this formulation was reworked to leverage the scheduling toolkit of the CPLEX optimization software package.

To illustrate the concept, let us first consider the constraint program for an operad to plan tasks without time and then add the time dimension\footnote{Simply increasingly dimensionality is not computationally wise--which was the point of exploring the CPLEX scheduling toolkit to address the time dimension--but this model still serves as a conceptual reference point.}.
Operad types are translated to boolean vectors $m_j$--whose entries capture individual agents at discrete coordination locations.  Parallel composition of primitive operations is expressed with boolean vectors $\Sigma_j$ indexed over primitive tasks for specific agents.
Type vectors $m_j$ indicate the coordination location each agent with value one; operation vectors $\Sigma_j$ indicate which tasks are planned in parallel.  

Assuming an operation with task vector $\Sigma_j$ and source vector $m_j$,
the target %
is $m_{j + 1} = m_j + M \Sigma_{j} $
where $M$ describes the relationship between source and target for primitive tasks.
Rows of $M$ correspond to primitive tasks while columns correspond to individual agents. 
The target to source constraint for single step of in-series composition is $
  m_{j+1} \ge M^{s} \Sigma_{j+1}$
where $M^{s}$ has rows that give requirements for each primitive task.  Here the LHS describes the target and the RHS describes the source.  The inequality appears to allow for implicit identities for agents without tasking--e.g.\ if $\Sigma_j$ is a zero vector, then $m_{j+1} = m_j$. This constraint prevents an individual agent from being assigned conflicting tasks or `teleporting' to begin a task.

 As seen in Fig.\ \ref{fig:PTconstraint}, additional agents: (1)  enable  more primitive tasks, indexed by Petri net transitions (top two rows); and (2) expand the type vector/matrix column dimension to account for new agent-location pairs and increase the matrix row dimension to account for new tasks (bottom two rows).  
 For example, the first 4 rows of $M(\HH \otimes \HH)$ correspond to the image of $\mathsf{\Lambda}(\HH) \times \mathsf{\Lambda}(\HH)$ in  $\mathsf{\Lambda}(\HH \otimes \HH)$.  The last row corresponds to new task, $\tau_4$, for the available pair of UH-60s. 
During implementation, the constraints can be declared task-by-task/row-by-row to sparsely couple the involved agents.  
 Once a limit on the number of steps of in series composition is set--i.e.\ a bound for the index $j$ is given--a finite constraint program is determined.
 
Time is readily modeled discretely with tasks given integer durations.  This corresponds to a more detailed network model, $\mathsf{\Lambda}_t$, whose types include a discrete time index;
see Fig.\ \ref{fig:exTaskOp} for example operations. 
Under these assumptions, one simply replaces the abstract steps of in series composition with a time index and decomposes $M$ and $\Sigma_j$ by the duration $d$ of primitive tasks:
\[
  m_t + \sum_{d=1}^{d_{\max}} M_d \Sigma_{t-d,d} = m_{t + 1}; ~~~~
  m_{t+1} \ge  \sum_{d=1}^{d_{\max}}M^{s}_{d} \Sigma_{t+1,d}
\]
so that $\Sigma_{t,d}$ describes tasks beginning at time $t$; the inequality allows for `waiting' operations. 
One can also model tasks more coarsely--with
$\mathsf{\Lambda}_{\bullet} \maps \NN(C) \to \Cat$--to construct an operad to task counts of agents without individual identity.
Then, the type vectors $m_j$ (resp., operation vectors $\Sigma_{j}$) have integer entries to express agent counts (resp., counts of planned tasks) with corresponding reductions in dimensionality.    
These three levels of network models 
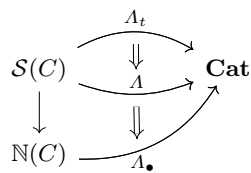
\[ \begin{tikzcd}
     \S(C) 
    \arrow[rr, "\mathsf{\Lambda}_t"{name=F}, bend left=30]
    \arrow[rr, ""{name=F, below}, bend left=30, pos=0.5]
    \arrow[rr, "\mathsf{\Lambda}"{name=F', above}, bend right=20]
    \arrow[rr, ""{name=F'', below}, bend right=20]
    \arrow[Rightarrow, from = F, to = F']
    \arrow[d]
    &  &\Cat 
    \\
    \N(C)
    \arrow[urr, "\mathsf{\Lambda}_{\bullet}"{below}, bend right=30, pos=0.39]
    \arrow[urr, ""{name=F''', above}, bend right=30, pos=0.35]
    \arrow[Rightarrow, from = F'', to = F''', swap]
\end{tikzcd}
\]
naturally induce morphisms of network operads\footnote{Strictly speaking, the coarsest (lowest) level is not network model; its domain is a free commutative monoidal category. Nevertheless, a completely analogous construction produces a typed operad fitting into this  diagram.} \cite[6.18]{NetworkModels} and encode mappings of syntactic variables that preserve feasibility.
In particular, the top two levels describe a precise mapping from task scheduling (highest) to task planning (middle). 
The lowest level $\mathsf{\Lambda}_{\bullet}$ forgets the individual identity of agents, providing a coarser level for planning. 

This very simple idea of enforcing type matching constraints is inherently natural\footnote{I.e., operad morphisms push forward feasible assignments variables in the domain to feasible assignments in the codomain.}.  However, further research is needed to determine if this natural hierarchical structure can be exploited by algorithms--e.g.\ by branching over pre-images of solutions to coarser levels--perhaps for domains were operational constraints coming from algebras are merely a nuisance, as opposed to being a central 
challenge for SAR planning.  For instance, a precise meta-model for planning and scheduling provides a common jumping off point to apply algorithms from those two disciplines.

{\bf Applying the cookbook: algebras.}  Because the operad template defines generating operations, specifying  algebras involves: (1)
capturing the salient features of each agent type as its internal state; and (2) specifying how these states update under generating morphisms--including, for operads with time, the implicit 
    `waiting' operations.   
For the SAR tasking problem, the salient features are fuel level and cumulative probability of survival throughout the mission. Typical primitive operations will not increase these values; fuel is expended or some risk is incurred. The notable exception is refueling operations which return the fuel level of the receiver to maximum.     
By specifying the non-increasing rate for each agent--location pair, the action of `waiting' operations are specified.  In practice, these data are derivable from environmental data for a scenario so that end users can manipulate them indirectly.

{\bf Operational constraints from algebras.}
  Salient features of each agent type are captured as auxiliary variables determined by syntactic decision variables.  The values of algebra variables are constrained of update equations--e.g.\ 
  to update fuel levels for agents with $
  \max(f_j + F \Sigma_{j}, f_{\max} )   = f_{j + 1}$,
where $f_{\max}$ specifies max fuel capacities.
 Having expressed the semantics for generating operations, one can enforce additional operational constraints--e.g.\ safe fuel levels: $
 f_{j + 1} \ge f_{\min}.$

{\bf Extending the domain of application.}
As noted above, this sparse declaration of a tasking domain is readily extended--e.g.\ to add a new atomic type or new ways for agents to coordinate.
Syntactically, this is amounts to new elements of $C$ or transitions to define primitive tasks.  Semantics must capture the impact of primitive operations on state, which can be roughly estimated initially and later refined.  
This flexibility is especially useful for rapid prototyping of  `what if' options for asset types and behaviors, as the wildfire SAR tasking problem illustrates.

Suppose, for example, that we wanted to model a joint SAR and fire fighting problem.
Both domains are naturally expressed with network operads to task behavior. Even if the specification formats were independently developed: (1) each format must encode the essential combinatorial data for each domain; and (2) category theory provides a method to integrate domain data: construct a pushout.
Analogous to taking the union of two sets along a common intersection, one identifies the part of the problem common to both domains--e.g.\ MAFFS-equipped HC-130s and their associated tasks appearing in both domains--to construct a cross-domain model
 
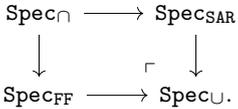
\[ \begin{tikzcd}
     \mathtt{Spec_\cap}
     \arrow[dr, phantom, "\ulcorner", very near end]
    \arrow[r]
    \arrow[d] 
  &\mathtt{Spec_{SAR}} \arrow[d]
    \\
    \mathtt{Spec_{FF}} \arrow[r] & \mathtt{Spec_\cup}.
\end{tikzcd}
\]
The arrows in this diagram account for translating the file format for the overlap into each domain-specific format and choosing a specific output format for cross-domain data.

On the other hand, suppose that the machine readable representation of each domain was tightly coupled to algorithms--e.g.\ mathematical programming for SAR and planning framework for fire fighting.
There is no artifact suitable for integrating these domains since expression was prematurely optimized.
We describe a general workflow to separate specification from representation and exploitable data structures and algorithms in \ref{sec:sepcon}\ref{subsec:opProg}. 

\subsection{Other examples of automated synthesis}
\label{subsec:autoOther}
Though network templates facilitate exploration from atoms, how to explore valid designs is a largely distinct concern from defining the space of designs, as discussed in \ref{sec:intro}. 

{\bf Novel search strategies via substitution}
For example, in the DARPA Fundamentals of Design (FUN Design) program, composition of designs employed a genetic algorithm (GA).
 FUN Design focused on generating novel conceptual designs for mechanical systems--e.g.\ catapults to launch a projectile. 
Formulating this problem with network operads followed the cookbook approach: there were atomic types of mechanical components and basic operations to link them.

The operad-based representation provided guarantees of design feasibility and informed how to generalize the GA implementation details.
Specifically, composition for atomic algebra elements defined genetic data; crossover produced child data to compose from atoms; and mutation  modified parameters of input algebra elements.
Crafting a crossover step is typically handled case-by-case while this strategy generalizes to other problems that mix combinatorial and continuously varying data, provided this data is packaged as an operad acting on an algebra.
  Guarantees of feasibility dramatically reduced the number unfit offspring evaluated by simulation against multiple fitness metrics.  Moreover, computational gains from feasibility guarantees increase as the design population becomes more combinatorially complex. 

{\bf Integrated structure and behavior}
Large classes of engineering problems compose components to form an `optimized' network--e.g.\ in chemical process synthesis, supply chains, and water purification networks \cite{khor2012superstructure, neiro2003supply, raman1992integration, yeomans1999systematic}.  Given a set of inputs, outputs and available operations (process equipment with input and output specification), the goal is to identify the optimal State Equipment Networks (SEN) for behavioral flows of materials and energy.
A given production target for outputs is evaluated in terms of multiple objectives such as environmental impact and cost.
For example, the chemical industry considers the supply chain, production and distribution network problem \cite{yeomans1999systematic} systematically as three superstructure optimization problems that can be composed to optimize enterprise level, multi-subsystem structures. 
Each sub-network structure is further optimized for low cost and other metrics including waste, environmental impact and energy costs. 
The operadic paradigm would provide a lens to generalize and refine existing techniques to jointly explore structure and behavior. 

  CASCADE prototyped integrated composition of structure and behavior for distributed logistics applications.  Here an explicit resupply plan to task agents was desired.  Structural composition was needed to account for the resupply capacity for heterogeneous delivery vehicles and the positioning of distributed resupply depots. %
  Probabilistic models estimated steady state resupply capacities of delivery fleet mixes to serve estimates of demand.  First, positioning resupply locations applied hill climbing to minimize the expected disruption of delivery routes when returning to and departing from resupply locations. Second, this disruption estimate was used to adjust the resupply capacity estimate of each delivery asset type. Third, promising designs where evaluated using a heuristic task planning algorithm.     
  At each stage, algorithms focused on finding satisficing solutions which allowed broad and rapid explorations of the design and tasking search space.
  
{\bf Synthesis with applied operads and categories.}
Research activity to apply operads and monoidal categories to automated design synthesis is increasing.  Wiring diagrams have been applied to automate protein design  \cite{Matriarch, SpiProtein} and collaborative design \cite[Ch.\ 4]{Seven} of physical systems employing practical semantic models and algorithms \cite{Co-design, Co-designUn, CoDesignScale, CoDesignIntel}.  Software tools are increasingly focused on scaling up computation, e.g.\  \cite{DisCoPy, Catlab, PyZX}, as opposed to software to augment human calculation, as in \cite{Glob, Quantomatic, HomotopyIO},
and managing complex domains with commercial-grade tools \cite{BSW, SpivakWis, CompPow, MultiManu}.
Recent work to optimize quantum circuits \cite{OptQuant, ReduceQuant} leverages such developments. The use of wiring diagrams to improve computational efficiency via normal forms is explored in \cite{NormWire}.

In the next section, we discuss research directions to develop the meta-modeling potential of applied operads to: (1) decompose a problem within a semantic model to divide and conquer; and (2) move between models to fill in details from coarse descriptions.
We also discuss how the flow of representations used for SAR--network template, operad model of composition, exploitation data structures and algorithms--could be systematized into a reusable software framework.

\section{Toward practical automated analysis and synthesis}
\label{sec:sepcon}

In this section, we describe lessons learned from practical experiences with applying operads to automated synthesis.
We frame separation of concerns in the language of operads to describe strategies to work around issues raised by this experience. 
This gives not only a clean formulation of separation but also a principled means to integrate and exploit concerns.

\subsection{Lessons from automated synthesis}
\label{subsec:autolessons}
The direct, network template approach facilitates correct and transparent modeling for complex tasking problems. 
However, computational tractability is limited to small problems--relative to the demands of applications.
More research is needed to develop efficient algorithms that break up the search into manageable parts, leveraging the power of operads to separate concerns.

Under CASCADE, we experimented with the CPLEX scheduling toolkit to informally model across levels of abstraction and exploit domain specific information.
In particular, generating options to plan, but not schedule, key maneuvers with traditional routing algorithms helped factor the problem effectively. 
These applied experiments were not systematized into a formal meta-modeling approach, although our prototype results were promising.
Specification of these levels--as in Sec.\ \ref{sec:cookbook}--and controlling the navigation of levels using domain-specifics would be ideal. 

The FUN DESIGN genetic algorithm approach illustrates the potential operads have to: (1) generalize case-by-case methods\footnote{In fact, applying genetic algorithms to explore network structures was inspired by the success of NeuroEvolution of
Augmenting Topologies (NEAT) \cite{NEAT} to generate novel neural network architectures.}; (2) separate concerns, in this case by leveraging the operad syntax for combinatorial crossover and algebra parameters for continuous mutation; and (3) guarantee correctness as complexity grows.
Distributed logistics applications in CASCADE show the flexibility afforded by multiple stage exploration for more efficient search.

\subsection{Formal separation of concerns}
\label{subsec:formalsepcon}

We begin by distinguishing {\it focus} from {\it filter}, which are two ways operads separate.  
Focus selects \emph{what} we look at, while filter captures \emph{how} we look at it.  These are questions of syntax and semantics, respectively. To be useful, the \emph{what} of our focus must align with the \emph{how} of the filter.

Separation of focus occurs within the syntax operad of system maps. In  \ref{sec:operads}\ref{subsec:tree_view}, four trees correspond to different views on the same system. We can zoom into one part of the system while leaving other portions black-boxed at a high level.
Varying the target type of an operation changes the scope for system composition, such as restricting attention to a subsystem.

Filtering, on the other hand, is semantic; we choose which salient features to model and which to suppress, controlled by the semantic context used to `implement' the operations.
As described in \ref{sec:anal}\ref{subsec:analyis}, the default semantic context is $\Set$ where: (1) each type in the operad is mapped to a set of possible instances for that type; and (2) each operation is mapped to a function to compose instances. 
Instances or algebra elements for the sailboat problem (Sec.~\ref{sec:cookbook}) describe the key features of structural system designs.  For SAR tasking (Sec.~\ref{sec:auto}), mission plan instances track the key internal states of agents--notably fuel and risk--throughout its execution.
 Section \ref{sec:anal} illustrates alternative semantic contexts as such  
probability $\Prob$ or relations between sets $\Rel$. 

Focus and filter come together to solve particular problems.  The analysis of the LSI system in Sec.\ \ref{sec:anal} tightly focuses the syntax operad $\WW$ to include only the types and operations from Fig.\ \ref{fig:decomps}.
Formally, this is accomplished by considering the image of the generating types and operations in the operad of port-graphs \cite[3]{LSIPaper}. 
This tight focus means semantics need only be defined for LSI components.  In each SAR tasking  problem of Sec.\ \ref{sec:auto}, an initial, source configuration of agent types is given, narrowing the focus of each problem.  The SAR focus is much broader because an operation to define the mission plan must be constructed. 
Even so, semantics filter down to just the key features of the problem and how to update them when generating operations act.     

Functorial semantics, as realized by an operad algebra $\Alg \maps \OO \to \Sem$, helps factor the overall problem model to facilitate its construction and exploitation.
For example, we can construct the probabilistic failure model in Table~\ref{tab:fail} by normalizing historical failures.  First we limit focus from all port-graphs $\PP$ to $\WW$
then semantics for counts in $\NN^+$, an operad of counts and sums, are normalized to obtain probabilities in $\Prob$:
 \[ \begin{tikzcd}
     \WW 
    \arrow[r, ""{name=F}]
    \arrow[dr, "\Alg"{name=F}, dashed]
    \arrow[d] 
   &\NN^+ \arrow[d]
    \\
    \PP & \Prob.
\end{tikzcd}
\]

The power to focus and filter is amplified because we are not limited by a single choice of how to filter.  In addition to {\it limiting} focus with the source of an operad algebra, we can {\it simplify} filters.  Such natural transformations between functors are  `filters of filters' that align different compositional models precisely--e.g.\ requirements over state (\ref{sec:anal}\ref{subsec:LSI}) or timed scheduling over two levels of planning (\ref{sec:auto}\ref{subsec:autoTask}). 
In this first case the syntax operad $\WW$ stays the same and semantics are linked by an algebra homomorphism (\ref{sec:operads}\ref{subsec:mapmaker}).  In the second case, both the operad and algebra must change 
to determine simpler semantics--e.g.\ to neglect the impact of waiting operations, which bound performance.
Such precision supports automation to explore design space across semantic models and aligns the ability to focus within each model. By working backward relative to the construction process,  we can lift partial solutions to gradually increase model fidelity--e.g.\ exploring schedules over effective plans.  This gives a foundation for lazy evaluation during deep exploration of design space, which we revisit in \ref{sec:sepcon}\ref{subsec:opProg}.

For a simple but rich example of these concepts working together, consider the functional decomposition $\varphiOp(\lambdaOp,\tauOp)$ in Fig.~\ref{fig:decomps}. We could model the length system $\lambdaOp$ using rigid-body dynamics, the temperature system $\tauOp$ as a lumped-element model, and super-system $\varphiOp$ as a computation (Edl\'en equation) that corrects the observed $\fringe$ count based on the measured temperature:
\begin{equation}
    \xymatrix{
\{\lambdaOp\} \ar@{>->}[d] \ar[r] & \Net_{\mathrm{mech}} \ar[rr]^-{\Rigid} && \Dyn \ar[d]^{\Impl}\\
\WW \ar[r] \ar@{}[rrrd]|{\displaystyle\Uparrow} \ar@{}[rrru]|{\displaystyle\Downarrow} & \Net_{\mathrm{comp}} \ar[rr]^-{\Edlen} && \Type\\
\{\tauOp\} \ar@{>->}[u] \ar[r] & \Net_{\mathrm{therm}} \ar[rr]^-{\Lump} && \Dyn \ar[u]_{\Impl} \\
}
\label{eqn:SB}
\end{equation}
The upper and lower paths construct implementations of dynamical models based on the aforementioned formalisms. The center path implements a correction on the data stream coming from the interferometer, based on a stream of temperature data. The 
two natural transformations
indicate extraction of one representation, a stream of state values, from the implementation of the dynamical models. Composition in $\WW$ then constructs the two data streams and applies the correction.

A key strength of the operadic paradigm is its genericity: the same principles of model construction, integration and exploitation developed for measurement and SAR apply to all kinds of systems. In principle, we could use the same tools and methodology to assemble transistors into circuits, unit processes into chemical factories and genes into genomes. The syntax and semantics change with the application, but the functorial perspective remains the same. 
For the rest of this section, we describe research directions to realize such a general purpose vision to
decompose a complex design problem into subproblems and support rapid, broad exploration of design space.

\subsection{Recent advancements, future prospects and limits}  
\label{subsec:advancementsAndLimits}

{\bf Progress driven by applications.} Section \ref{sec:cookbook} describes how cookbook-style approaches enable practitioners to put operads to work.
Generative data define a domain and compositionality combines it into operads and algebras to separate concerns. 
Network operads \cite{NetworkModels,  NMPetri, Moeller}
were developed {\it in response to the demands of applications}  
to construct operads from generative data.
Section \ref{sec:anal} describes rich design analysis by leveraging multiple decompositions of complex systems and working across levels of abstraction.  
Focusing on a specific applied problem--the LSI at NIST--provided further opportunities for analysis since {\it model semantics need only be defined for the problem at hand}; see also Eq. \ref{eqn:SB}.   
Progress in streamlining  automated synthesis from building blocks is recounted in Sec.\ \ref{sec:auto}
where the domain drives coordination requirements to task behavior. 

{\bf Prospects.} If interactions between systems are well-understood  
(specification) and can be usefully modeled by compositional semantics (analysis), then automated design synthesis leveraging separation for scalability becomes possible. 
For instance, most references from the end of Sec.\ \ref{sec:anal} correspond to domains that are studied with diagrams that indicate interactions and have associated compositional models.
This allows intricate interactions to be modeled--compare, e.g.\ classical \cite{EDAVLSI} vs. quantum \cite{QM, OptQuant, ReduceQuant} computing--while unlocking separation of concerns.
Cookbook and focused approaches guide practitioners to seek out the minimal data needed for a domain problem--as in the examples presented--but operads for design {\it requires} compositional models. 

{\bf Limitations.} 
We note three issues limiting when operad apply: (1) key interactions among systems and components are {\it inputs};
(2) not all design problems become tractable via decomposition and hierarchy; and
(3) there is no guarantee of compositional semantics to exploit.
 For instance, though the interactions for the $n$-body problem are understood (1), this does not
 lend itself to decomposition (2) or exploitable compositional semantics (3).
 Whitney \cite{Whitney} notes that integral mechanical system design must address safety issues at high power levels due to challenging, long-range interactions.  
 Some aspects of mechanical system design may yield to operad analysis--e.g., bond graphs \cite{CoyaThesis} or other sufficiently ``diagrammatic'' models--but others may not.
 Both examples illustrate how overnumerous or long range interaction can lead to (2).  Operads can work at the system rather than component level if system properties can be extracted into compositional models.
 However, operads do not provide a means to extract such properties or understand problems that are truly inseparable theoretically or practically.

\subsection{Research directions for applied operads}  
\label{subsec:research}
We now briefly overview research directions toward automated analysis and synthesis.

{\bf Operad-based decomposition and adaptation.} 
Decomposition, ways a complex operation can be broken down into simpler operations, is a dual concept to the composition of operations.
Any subsystem designed by a simpler operation can be adapted: precisely which operations can be substituted is known, providing a general perspective to craft algorithms.  
To be practical, the analytic questions of \emph{how} to decompose and \emph{when} to adapt subsystems must be answered.

One research direction applies the lens of operad composition to abstract and generalize existing algorithms that exploit decomposition--e.g.\ to: (1) generalize superstructure optimization techniques discussed in \ref{sec:auto}\ref{subsec:autoOther}; extend
(2) extend the crossover and mutation steps for the  FUN DESIGN work \ref{subsec:autoStruct}, which are global in the sense that they manipulate full designs, to local steps which adapt parts of a design, perhaps driven by analysis to re-work specific subsystems; and (3) 
explore routing as a proxy for tasking planning, analyzing foundational algorithms like Ford-Fulkerson \cite{FF} and decomposition techniques such as contraction hierarchies \cite{CH}.
An intriguing, but speculative, avenue is to attempt to learn how to decompose a system or select subsystems to adapt in a data-driven way, so that the operad syntax constrains otherwise lightly supervised learning.
A theoretical direction is to seriously consider the dual role of decomposition, analogous to Hopf and Frobenius algebra \cite{Dusko}, and attempt to gain deeper understanding of the interplay of composition and decomposition, eventually distilling any results into algorithms\footnote{For example, Bellman's principle of optimality is \emph{decompositional}--i.e.\ parts of an optimal solution are optimal. 
}. 

{\bf Multiple levels of modeling.}
The LSI example  shows how a system model can be analysed to address different considerations. 
This sets the stage to adapt a design--e.g.\ bolster functional risk points and improve control in back and forth fashion--until both considerations are acceptable. 
Applied demonstrations for SAR tasking suggest a multi-level framework: 
(1) encoding operational concepts; 
(2) planning options for key maneuvers; and
(3) multistage planning and scheduling to support these maneuvers. 

 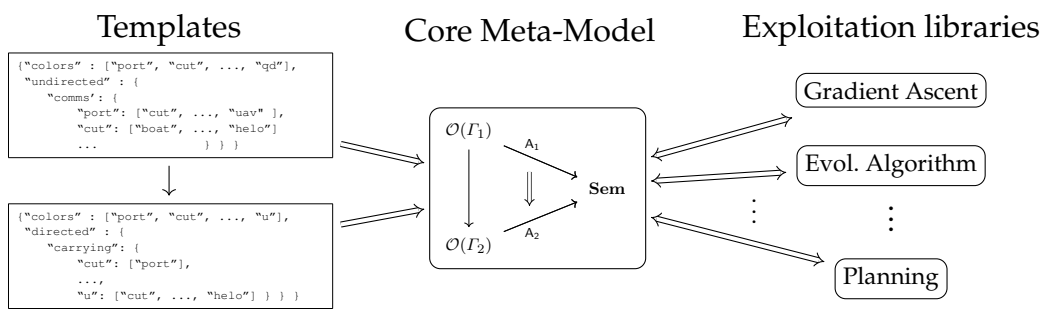
\begin{figure}[b]
\begin{center}
\begin{tikzpicture}
	\begin{pgfonlayer}{nodelayer}
		\node [style=none]
			(Temps) at (-5, -0.2) {\large Templates};
 		\node [style=none, outer sep =4pt]
 			(Fine) at (-5, -1.2) {
 		\scalebox{0.5}{
		\fbox{\parbox{0.6\textwidth}{
        \texttt{\{``colors'' : [``port'', ``cut'', \ldots , ``qd''], \\
         \hspace*{2mm}``undirected'' : \{ \\
         \hspace*{6mm} ``comms': \{ \\
         \hspace*{6mm} \hspace*{6mm} ``port'': [``cut'', \ldots, ``uav" ], \\
        \hspace*{6mm} \hspace*{6mm} ``cut'': [``boat'', \ldots , ``helo'']
        \hspace*{6mm} \hspace*{6mm}  \ldots \hspace*{12mm} \hspace*{12mm}  \} \} \} 
        }}}}
 			};
 		\node [style=none, outer sep =4pt]
 			(Coarse) at (-5, -3.2) {
 		\scalebox{0.5}{
		\fbox{\parbox{0.6\textwidth}{
        \texttt{\{``colors'' : [``port'', ``cut'', \ldots , ``u''], \\
         \hspace*{2mm}``directed'' : \{ \\
         \hspace*{6mm} ``carrying'': \{ \\
         \hspace*{6mm} \hspace*{6mm} ``cut'': [``port''], \\
        \hspace*{6mm} \hspace*{6mm}  \ldots, \\
        \hspace*{6mm} \hspace*{6mm}  ``u'': [``cut'', \ldots , ``helo''] \} \} \} 
        }}}}
 			};
		\node [style=none]
			(KR) at (-0.25, -0.2) {\large Core Meta-Model};
			\node [draw, rounded corners, outer sep =2pt]
			(Comm) at (-0.15, -2.3) {
	\adjustbox{scale=0.7}{%
 	\begin{tikzcd}
     \OO(\mathsf{\Gamma}_1)
    \arrow[drr, "\mathsf{A_1}"{above}, pos=0.39]
    \arrow[drr, ""{name=F, below}, pos=0.35]
    \arrow[dd]  & &
    \\
    &  &  \Sem
    \\
    \OO(\mathsf{\Gamma}_2)
    \arrow[urr, "\mathsf{A_2}"{below}, pos=0.39]
    \arrow[urr, ""{name=F', above}, pos=0.35]
    \arrow[Rightarrow, from = F, to = F', swap]
\end{tikzcd}
}
};
			\node [style=none]
			(KE) at (4.5, -0.2) {\large Exploitation  libraries};
		    \node [draw, rounded corners, outer sep =4pt]
			(Grad) at (4.5, -1) {\small Gradient Ascent};
			\node [draw, rounded corners, outer sep =4pt]
			(Evol) at (4.5, -2) {\small Evol.\  Algorithm};
			\node [style=none]
			(dotsBig) at (4.5, -2.6) {\large \vdots};
			\node [draw, rounded corners, outer sep =4pt]
			(Plan) at (4.5, -3.5) {Planning};
			\node [style=none]
			(dotsSmall) at (2.7, -2.5) {\vdots};
	\end{pgfonlayer}
	\begin{pgfonlayer}{edgelayer}
	%
	%
	\draw [->]  (Fine) to node[midway, left] {} (Coarse);
	\draw [-implies, style=black,
		double equal sign distance]  (Fine) to node[midway, above] { } (Comm);
	\draw [-implies, style=black,
		double equal sign distance]  (Coarse) to node[midway, above] { } (Comm);
	\draw [implies-implies, style=black,
		double equal sign distance]  (Comm) to node[midway, above] { } (Grad);
	\draw [implies-implies, style=black,
		double equal sign distance]  (Comm) to node[midway, above] { } (Evol);
	\draw [implies-implies, style=black,
		double equal sign distance]  (Comm) to node[midway, above] { } (Plan);
	\end{pgfonlayer}
\end{tikzpicture}
\end{center}
\vspace*{-12pt}
 \caption[caption]{A software framework to leverage a meta-model: templates define each level and how to move between, libraries exploit each level, and core meta-model facilitates control across levels.}
 \vspace*{-24pt}
 \label{fig:system}
 \end{figure}

{\bf Unifying top-down and bottom-up points of view.}  We have laid out the analytic--exemplified by wiring diagrams--and synthetic--exemplified by network operads--points of view for complex systems.  
Even if the goal is practical automated synthesis, scalability issues promote analytic decomposition and abstraction to efficiently reason toward satisficing solutions.
Two approaches to unification   
include: (1) create a combined syntax for analysis and synthesis, a `super operad' combining both features; (2) act by an analytic operad on the synthetic syntax, extending composition of operations.
While the former approach is arguably more unified, the later more clearly separates analysis and synthesis and may provide a constructive approach to former.

\subsection{Functorial programming with operads}
\label{subsec:opProg}

 At this point, experience implementing operads for design suggests a software framework. 
 While conceptually simple, this sketch helps clarify the practical role of a precise meta-model.

Rather than working directly with operads to form a core meta-modeling language, cf.\ \cite{MultiMod}, a workflow akin to popular frameworks for JavaScript  development would put developers in the drivers seat: adopters focus on controlling the flow of data and contribute to an ecosystem of libraries for lower-level data processing.
Achieving this requires work before and after the meta-model.  
First, transferable methods get an applied problem into operads (Fig.\ \ref{fig:system}, left).  As in Section \ref{sec:cookbook}, this data constructs operads and algebras to form the core meta-model. Core data feeds explicitly exploitable data structures and algorithms to analyze (Sec.~\ref{sec:anal}) and automatically construct (Sec.~\ref{sec:auto}) complex systems (Fig.\ \ref{fig:system}, right).  
On far the left, end user tools convert intent to domain inputs.
Rightmost, libraries access exploitation data structures and algorithms, including those exploiting the syntax and semantics separation 
 or substitution and adaptation.  At the center, the core meta-model guarantees that the scruffier ends of the framework exposed to end users and developers are correctly aligned and coherently navigated.
 
This framework provides significant opportunities to separate concerns
compared to other approaches.  Foremost, the core model separates syntax from semantics.
As noted in \ref{sec:intro}, applied methods tend to conflate syntax and semantics.
 For instance, aggregate programming \cite{AggProg} provides: 1) semantics for networked components with spatial and temporal extent; and (2) interactions are proximity-based.  The former feature is powerful but limiting: by choosing a single kind of semantics, modeling is wedded to the scales it abstracts well. 
 The individual component scale is not modeled, even syntactically, which would  complicate any attempt to align with other models.
The latter precludes syntactic declaration of interactions--e.g.\ to construct architectures not purely based on proximity--and the absolute clarity about what can be put together provided by the operad syntax. 
Relative to computational efforts to apply operads or monoidal categories, e.g.\ \cite{DisCoPy, Catlab}, this sketch places greater emphasis on specification and exploitation: specification of a domain is possible without exposing the meta-model, algorithms searching within each model are treated as black boxes that produce valid designs. Separate specification greatly facilitates set up 
by experts in the domain, but not the meta-model.  Separate exploitation encourages importing existing data structures and algorithms to exploit each model.  
 
\subsection{Open problems}
 \label{subsec:Open}

The software framework just sketched separates out the issues of  practical specification, meta-modeling and fast data structures and algorithms.
We organize our discussion of open problems around concrete steps to advance these  issues. 
In our problem statements, ``multiple'' means at least three to assure demonstration of the genericity of the operadic paradigm.

{\bf Practical specification.} The overarching question is whether the minimal combinatorial data which can specify operads, their algebras and algebra homomorphisms in theory can be practically implemented in software.
We propose the following problems to advance the state-of-the-art for network template specification of operads described in Sec.\ \ref{sec:cookbook}:
\begin{enumerate}
    \item Demonstrate a specification software package for operad algebras for multiple domains.
    \item Develop specification software for algebra homomorphisms to demonstrate correctly aligned navigation between multiple models for a single domain.
    \item Develop and implement composition of specifications to combine multiple parts of a domain problem or integrate multiple domains.
\end{enumerate}
This last point is inline with the discussion of extending a domain in \ref{sec:auto}\ref{subsec:autoTask} and motivates a need to reconcile independently developed specification formats.   
\begin{enumerate}
\setcounter{enumi}{3}
    \item Demonstrate automatic translation across specification formats.
\end{enumerate}

{\bf Core meta-model.} 
As a practical matter, state-of-the-art examples exercise general principles of the paradigm but do not leverage general purpose software to encode the meta-model.
 \begin{enumerate}
    \setcounter{enumi}{4}
     \item Develop and demonstrate reusable middleware to explicitly encode multiple semantic models and maps between them which
     (a) takes inputs from specification packages; and (b) serves as a platform to navigate models.
    \end{enumerate}
We have seen rich examples of focused analysis with wiring diagrams in Sec.\ \ref{sec:anal} and automated composition from building blocks in Sec.\ \ref{sec:auto}.  Theoretically, there is the question of integrating the top-down and bottom-up perspectives: 
 \begin{enumerate}
    \setcounter{enumi}{5}
    \item Develop unified foundations to integrate: (a) analytic and synthetic styles of operads; and (b) composition with decomposition.
    \end{enumerate}
Potential starting points for these theoretical advancements are described in \ref{sec:sepcon}\ref{subsec:research}.
Developing understanding of limitations  overviewed in \ref{sec:sepcon}\ref{subsec:advancementsAndLimits} requires engagement with a range of applications:
 \begin{enumerate}
    \setcounter{enumi}{6}
    \item Investigate limits of operads for design to:
     (a) identify domains or specific aspects of domains lacking minimal data; (b) demonstrate the failure of compositionality for potentially useful semantics; and (c) characterize complexity barriers due to integrality.
    \end{enumerate}

{\bf Navigation of effective data structures and algorithms.}
Lastly, there is the question of whether coherent navigation of 
    models can be made practical.
    This requires explicit control of data across models and fast data structures and algorithms within specific models.
The general-purpose evolutionary algorithms discussed in \ref{sec:auto}\ref{subsec:autoOther} motivate:
    \begin{enumerate}
    \setcounter{enumi}{7}
    \item Develop reusable libraries that exploit (a) substitution of operations and instances to adapt designs and (b) separation of semantics from syntax.
    \end{enumerate}
SAR tasking experience and prototype explorations for distributed logistics illustrate the need to exploit moving \emph{across} models:  
    \begin{enumerate}
    \setcounter{enumi}{8}
    \item  Develop and demonstrate general purpose strategies to exploit separation across models via  hierarchical
    representation of model fidelity--e.g.\ example: (a) Structure over behavior; and (b) planning over scheduling.
     \item Quantify the impact of separation of concerns on: (a) computational complexity; and (b) practical computation time.
    \end{enumerate}
For this last point, \emph{isolating} the impact of each way to separate  concerns is of particular interest to lay groundwork to systematically analyze complex domain problems.  
Finally, there is question of demonstrating an end-to-end system to exploit the operadic,  meta-modeling paradigm.
    \begin{enumerate}
    \setcounter{enumi}{11}
    \item Demonstrate systematic, high-level control of iteration,
    substitution  and moving across multiple models to solve a complex domain problem.
    \item Develop high-level control framework--similar to JavaScript frameworks for UI--or programming language--similar to probabilistic programming--to systematically control iteration,
    substitution  and movement across multiple models.
    \end{enumerate}

\section{Conclusion}

 Operads provide a powerful meta-language to unite complementary system models
 within a single framework. 
 They express multiple options for decomposition and hierarchy for complex  designs, both within and across models.
 Diverse concerns needed to solve the full design problem are coherently separated by functorial semantics, maintaining compositionality of subsystems.  
Each semantic model can trade-off precision and accuracy to achieve an elegant abstraction, while algorithms exploit the specifics of each model to analyze and synthesize designs.

The basic moves of iteration, substitution and moving across multiple models form a rich framework to explore design space.
The trade-off is that the technical infrastructure needed to fully exploit this paradigm is daunting.
Recent progress has lowered barriers to specify domain models and streamline tool chains to automatically synthesize designs from basic building blocks.
Key parts of relevant theory and its implementation in software have been prototyped for example applications.
Further research is needed to integrate advancements in automatic specification and synthesis with the analytic power of operads to separate concerns.
To help focus efforts, we described research directions and proposed some concrete open problems.

\enlargethispage{20pt}

\ethics{This article does not present research with ethical considerations.}

\dataccess{This article has no additional data.}

\aucontribute{All authors contributed to the development of the review, its text and provided feedback and detailed edits on the work as a whole.  JF coordinated the effort and led the writing of Sec.\ 6.  SB led Sec.\ 5 and co-led Sec.\ 2, 3 and 4 with JF.  ES focused on connections to applications while JD focused on assuring accessible mathematics discussion.}

\competing{We declare that we have no competing interests.}

\funding{JF and JD were supported by the DARPA Complex Adaptive System Composition and Design Environment (CASCADE) project under Contract No. N66001-16-C-4048.}

\ack{The authors thank John Baez, Tony Falcone, Ben Long, Tom Mifflin, John Paschkewitz, Ram Sriram  and Blake Pollard for helpful discussions and two anonymous referees for comments that significantly improved the presentation. }

\disclaimer{{Official contribution of the National Institute of Standards and Technology;} not subject to copyright in the United States. Any commercial equipment, instruments, or materials identified in this paper are used to specify the experimental procedure adequately. Such identification is not intended to imply recommendation or endorsement by the National Institute of Standards and Technology, nor is it intended to imply that the materials or equipment identified are necessarily the best available for the purpose.}


\end{document}